\documentclass[eqsecnum]{iopart}

\usepackage{bm}
\usepackage{epsf}
\usepackage{epsfig}
\usepackage{epsf}
\def \be{\begin{equation}\fl}
\def \ee{\end{equation}}

\def \o{\omega}

\def \K{G^{K}}
\def \R{G^{R}}
\def \A{G^{A}}

\def \D{{\cal D}}
\def \C{{\cal C}}

\newcommand{\req}[1]{Eq.~(\ref{#1})}
\newcommand{\reqs}[1]{Eqs.~(\ref{#1})}
\newcommand{\rref}[1]{(\ref{#1})}

\newcommand{\vare}{\varepsilon}
\newcommand{\mls}{\delta_1}
\newcommand{\esc}{\gamma_{\rm esc}}
\newcommand{\Nl}{N_{\rm l}}
\newcommand{\Nr}{N_{\rm r}}
\newcommand{\Nch}{N_{\rm ch}}
\newcommand{\gesc}{\gamma_{\rm esc}}
\newcommand{\Iph}{I_{\rm ph}}
\newcommand{\To}{\tau_{\rm o}}
\newcommand{\varIp}{{\rm var}\Iph}
\newcommand{\Gm}{{\cal T}}

\begin{document}

\bibliographystyle{prsty}

\title[Quantum chaotic scattering in time-dependent external fields]
{Quantum chaotic
scattering in time-dependent external fields: random matrix approach}


\author{Maxim~G.~Vavilov}
\address{Department of Applied Physics, Yale University, New Haven, CT 06520}
\ead{mv44@cornell.edu}

\begin{abstract}
We review the random matrix description of
electron transport through open quantum dots, subject to
time-dependent perturbations. All characteristics of the
current linear in the bias can be expressed in terms of the
scattering matrix, calculated for a time-dependent Hamiltonian.
Assuming that the Hamiltonian belongs to a Gaussian
ensemble of random matrices, we investigate various
statistical properties of the direct current in the ensemble.
Particularly, even at zero bias  the time-dependent perturbation
induces current, called photovoltaic current. We discuss dependence
of the photovoltaic current and its noise on the frequency and
the strength of the perturbation.
We also describe
the effect of time-dependent perturbation on the weak localization
correction to the conductance and on conductance fluctuations.
\end{abstract}


\maketitle

\section{Introduction}

Quantum dot is a small disordered or irregularly shaped conductor,
connected to leads~\cite{marcus}, see Fig. 1. Exact values of the conductance
of a quantum dot are determined by electron wave functions in the
system and are hard to calculate exactly for arbitrary
configurations of the dot. Moreover,  the conductance changes
significantly even for tiny changes in the position of impurities
or the boundary of the dot. Due to extreme sensitivity of the conductance on
many parameters, the statistical description of the conductance  is
more appropriate~\cite{Stone85,A1,LS85,AS86,Webb,BeenakkerRMP,AK,H1,HFPMDH}.
The random fluctuations of the conductance
from sample to sample
of non-interacting systems are universal. The universality~\cite{AS,AOS}
means that the conductance statistics can be described by universal
functions, which are independent from the shape of the dot or the
details of the disordered potential. Particularly,
the variance of the conductance ${\rm var} g$ is of
the order $G_0^2$ and is nearly  independent from the sample geometry
($G_0=e^2/\pi\hbar$ is the quantum of conductance for spin degenerate
electrons). The other universal quantity is the weak localization correction to the
conductance, defined as the difference of average values of the conductance
over orthogonal (zero magnetic field) and unitary (strong magnetic
field) ensembles. The weak localization correction to the conductance
is also of the order of $G_0$~\cite{WL,Argaman95,IWZ90,BJS93a,BJS93b,BM94,JPB}.

A common description of electron transport through quantum dots
is based on the Landauer formalism~\cite{Landauer,FL81,Bsymm,SS88},
when the transport characteristics of the system are described
in terms of the scattering amplitudes between different
conducting channels in the leads.
There are several approaches for statistical description
of electron transport.
One approach is based on a diagram technique developed
for disordered bulk metals~\cite{AGD},
when the scattering amplitudes are represented in terms of
electron Green functions~\cite{LS85,RS,BSD,BS89}.

Alternative
approaches are based on the description of the system by random matrices,
when either an exact scattering matrix is replaced by a random
unitary matrix, or an exact Hamiltonian is replaced
by a random Hermitian matrix.
In the first case, the unitary matrix is taken
from Dyson's circular ensemble of uniformly
distributed random matrices~\cite{BlSm88,BM94,JPB}.
In the Hamiltonian approach, the Hermitian
matrix belongs to an ensemble of random matrices~\cite{mehta}
with the Gaussian distribution of its matrix
elements~\cite{VWZ,LW91}.
The equivalence for statistical description of electron transport
by  both random matrix approaches was shown in~\cite{LW91,Brouwer95,FS97}.

Although the random matrix approach is not based on microscopic
description of electron system, their correspondence to microscopic problem
has been proven for disordered metal grains~\cite{Efetov82,Efetov83,Efetovbook}.
The validity of such
random matrix description of chaotic ballistic systems was addressed
in~\cite{MK97,AASA,AOS}.

We imply the following realization of the system, see Fig.~1 a).
Negative voltages applied to the gates (black areas) confine
electrons to a small region (light gray), forming a quantum dot.
Electrons in the dot are connected to the electron reservoirs by
narrow leads. Electric current that flows through the dot can be
measured as a function of the voltage bias $V$ between the reservoirs
and the amplitudes of ac gate voltages $V_{1,2}(t)$. Particularly,
the current linear in bias $V$ is determined by the
conductance of the dot. Changing magnetic field or
shape of the dot one can obtain different realizations
of the quantum dot and experimentally study statistics of the
quantum corrections to the conductance.

The quantum corrections to the conductance are commonly
characterized by the weak localization
and the variance of conductance fluctuations.
As any other quantum
interference phenomena, they are very sensitive to inelastic processes,
commonly referred to as dephasing~\cite{AAKL}. A phenomenological description of
the effect of dephasing on electron transport through
open quantum dots was developed in Refs.~\cite{BM95,BB95,McCL98}.
The dephasing rate due to electron--electron
interaction in quantum dots was estimated in  \cite{SIA,AGKL}.
Another possible source of dephasing is a time-dependent
perturbation, such as a microwave radiation or periodic deformation.
In this case the Hamiltonian of the system can be considered as a time-dependent
random matrix~\cite{VAwl,VAcf,PBSmat}, and all transport quantities
can be calculated as a function of various parameters
(e.g. strength and frequency) of the time-dependent perturbation.
The scattering matrix description of the system subject to
time-dependent perturbation was developed in energy representation
by B\"uttiker, Thomas  and Pretrein in Refs.~\cite{BTP93,BTP94}.
In this case the scattering
matrix describes processes when electron scattering  between
different channels in the leads is accompanied by the change of
electron energy. Alternatively, the analysis of the effect
of time-dependent perturbation on the conductance  can be  carried out in
time representation, see~Refs.~\cite{VAwl,VAcf,PBSmat,WKcf,YKK}.
In general, both the weak localization correction to the
conductance and the variance of conductance fluctuations are
suppressed by time-dependent perturbation.  The suppression of
the quantum corrections to the conductance by microwave radiation
was observed experimentally in \cite{HFPMDH}.

Time-dependent perturbation of quantum dots not only
suppresses quantum corrections to the conductance, but also
produces electric current through the system even at zero bias.
This effect is related to the charge pumping, which occurs in
systems with large tunnel barriers~\cite{Th,Niu,AL91,Kouw91}.
If the conductance of the system is very small,
the electric current is quantized in units of $e\omega/2\pi$,
where $2\pi/\omega$ is the period of the pump. At finite
conductance, a countercurrent reduces the pumped current and thus
violate the quantization of electric current~\cite{AApump}. For an
open quantum dot, the countercurrent nearly  compensates
the pumped current and the current is no longer quantized.

In the low frequency limit, the magnitude of the pumped current is
determined entirely by the evolution of the system
in the parameter space, see Fig. 1 b),  under time
dependent perturbations~\cite{Brouwer98,SZA,SAA,BH01}.
As frequency increases, the parametric
description becomes insufficient and requires full analysis
of electron dynamics in time-dependent fields~\cite{VAA,PBSmat}.
The analysis of how the adiabatic description
breaks down at finite frequency can be also found in
Refs.~\cite{WWG,EWAL}. We note that  the charge pumping
through an open quantum dot is a manifestation
of the photovoltaic effect, which occurs in systems without
inversion center~\cite{Belinicher}.
The photovoltaic effect was previously  considered  by Falko and
Khmelnitskii~\cite{FK} in mesoscopic microjunctions and by
Kravtsov, Aronov and Yudson~\cite{KrAr93,KrY93} in normal metal
rings.

It turns out~\cite{VAA} that the photovoltaic current is sensitive
to the actual electron distribution function in the dot.
Time-dependent perturbations may broaden the distribution function,
resulting in heating. This broadening of the electron distribution
occurs as a result the electron diffusion in the energy space.
The effect of time-dependent perturbations on electron
distribution function becomes even more interesting in closed
systems, when the energy diffusion
acquires quantum interference corrections. The latter leads to a
dynamic localization~\cite{FGP} of the electrons in energy
space~\cite{BSK,BKcb}.

Photovoltaic current fluctuates not only with respect to
different realizations of the quantum dot, but also
for a given  realization due
to quantum and thermal fluctuations. Such fluctuations are called
current noise and are described by the fluctuations
of the charge transported through the dot in a certain number of
perturbation cycles.
The statistics of such charge fluctuations was studied in Refs.\
\cite{KApump,Lpump,AM,LL,IL,buttikernew} for
temperatures $T$ and pumping frequencies $\omega$
much smaller than the inverse dwell time $\esc$ (escape rate) of electrons
from the quantum dot. Particularly, Refs.~\cite{KApump,Lpump} addressed the full counting
statistics at temperatures $ T \ll \omega$ (we use $\hbar=1$
and the Boltsmann constant $k_{\rm B}=1$). The
mean square charge fluctuations for $ \omega,  T \ll \esc$ (but
for arbitrary relation between $ \omega $ and  $T$) were considered
in Ref.~\cite{buttikernew}. The variance of
the photovoltaic current
for arbitrary
relation between the temperature $T$, the frequency $\omega$,
the escape rate $\esc$
and the strength of the perturbation was calculated in Ref.~\cite{PVB}.

Experiments~\cite{Kvon,linke,SwitkesP} were performed to
detect the photovoltaic current in various mesoscopic
systems~\cite{Kvon,linke}, including open quantum dots~\cite{SwitkesP}
in the adiabatic regime. The observed magnetic field symmetry and
the amplitude of the current indicate that the measured current
was likely related to the ac
rectification~\cite{BrouwerR,DCMH,VDCM}. A more detailed
analysis of the zero-bias current in different regimes of
microwave radiation shows that in some instances the photovoltaic
current, and not the rectification current, was observed~\cite{DCMH,VDCM}.

\begin{figure}
\begin{center}
\epsfxsize=12cm \epsfbox{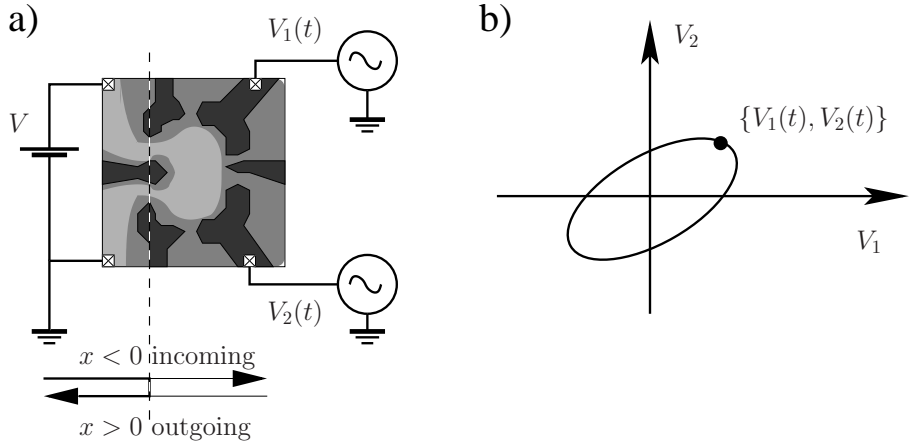}
\end{center}
\caption{\label{fig:0}
a) Schematic picture of the experimental setup. Light gray color
shows the region available for free electron motion, while the
dark gray color shows the region forbidden for electrons due to
electrostatic repulsion from the gates (shown in black) with
applied negative voltages. A finite bias $V$ is applied between
the upper and lower (in the text refereed to as left and right)
reservoirs. Oscillating voltages $V_{1,2}(t)$ applied to the gates
produce time-dependent perturbation of the electron system.
b) Contour plot represents time evolution of gate voltages
$V_{1,2}(t)$.
}
\end{figure}

In this review we focus on the random matrix
description of electron transport through
open quantum dots in the limit of the large number of open channels
$N_{\rm ch}$ connecting the dot to the leads. This condition
allows us to neglect the
electron--electron interaction that gives corrections of the
$1/N_{\rm ch}^2$ order,  see Ref. \cite{BA}. The same
condition permits the use of a diagrammatic technique, similar to
that described in \cite{AGD}, to calculate ensemble averaging.
We assume that the electron dynamics in the dot is fully chaotic
and disregard classical fluctuations of the
conductance~\cite{BSD}.
We emphasize that the random matrix description
is applicable for sufficiently small quantum dots, when the Thouless energy
$E_{\rm T}=1 / \tau_{\rm cross}$
is much greater than all other
energy scales of the problem, such as the frequency $\omega$
of the perturbation or the temperature $T$
($\tau_{\rm cross}$ is the electron crossing time of the dot).
Larger systems ($E_{\rm T} \ll \omega, \ T $) can be treated
by methods developed for bulk conductors~\cite{AAKL}, see e.g.
\cite{WKcf}. We note that the derivation of the results will be
performed within the Hamiltonian formalism, following
Refs.~\cite{VAwl,VAcf,VAA}, but the same results were derived
within scattering matrix formalism in \cite{PBSmat}.


\section{Scattering Matrix Formulation of Transport through Open Quantum Dots}
\label{sec:2}


\subsection{Model}
\label{sec:2.1}

The Hamiltonian of the system is 
\be
\label{1}
\hat {\cal H} =\hat {\cal H}_{\rm d}
+
\hat {\cal H}_{\rm ld}
+\hat {\cal H}_{\rm l}.
\ee
We choose the basis for electron wave functions in the dot,
so that the coupling of states in the dot  to states
in the leads  can be written as
\be
\label{7}
\hat H_{\rm ld}=\sum_{\alpha, n, k}\left( W_{n \alpha}\psi^\dag_\alpha
(k)
\psi_n+
{\rm H.c.}\right),
\quad
W_{n \alpha}=\cases{
\Gm,& if $n=\alpha\leq N_{\rm ch}$,\cr
0,& otherwise.}
\ee
Here  $\psi_n$ and $\psi_\alpha(k)$ are the annihilation operators
of electrons in the dot and the leads, respectively. Index
$n$ enumerates electron states in the dot: $n=1, \dots M$,
with $M\to\infty$. Index $\alpha$ labels channels in the leads, with
$1\leq \alpha\leq N_{\rm l}$ for the $N_{\rm l}$ channels in the left lead and
with $N_{\rm l}+1\leq \alpha \leq N_{\rm ch}$ for the $N_{\rm r}$ channels in
the right lead, $N_{\rm ch}=N_{\rm l}+N_{\rm r}$.
Coupling constants $\Gm$ are defined below in Eq.~(\ref{16z}).
The Hamiltonian
for electron states in the leads near the Fermi surface can be
linearized:
\be
\label{8}
\hat H_{\rm l}=v_{\rm F}\sum_{\alpha, k} k \psi^\dag_\alpha(k)\psi_\alpha(k),
\ee
where the continuous variable  $k$ denotes electron momenta in the leads,
$v_{\rm F}=(2\pi \nu)^{-1}$ is the Fermi velocity, and $\nu$ is the density of
states per channel at the Fermi surface.

Finally,  $\hat {\cal H}_{\rm d}$ is the Hamiltonian of the electrons in the
dot, determined by the $M\times M$ matrix $\hat H$ and the
electrostatic energy of $N$ electrons:
\be
{\cal H}_{\rm d}=
 \bm{\psi}^\dag \left[ \hat H + \sum_i (\hat V_i+\hat 1 Z_i)\varphi_i(t)\right]
 \bm{\psi} + E_{\rm c}N^2.
\label{Hd}
\ee
Matrix $\hat H$ describes the time independent part of the electron
Hamiltonian, and the time-dependent component of the Hamiltonian is
represented in terms of the traceless matrices $\hat V_i$  and the
diagonal matrix $\hat 1 Z_i $. In the setup shown in Fig.~1, the
time-dependent perturbation is generated by the gate voltages
$V_{1,2}(t)$.  The perturbation is linear in
small amplitude of oscillating voltages
$V_{1,2}(t)$, and the time evolution of the perturbation
characterized by the dimensionless functions $\varphi_i(t)\propto V_i(t)$.
The second term in ~\req{Hd} represents the largest in
$1/M$ contribution from the electron-electron interaction with
$E_{\rm c}$ being the charging energy of the dot, and
$N= \sum_{n}\psi^\dagger_n \psi_n$ being the operator of the electron number
in the dot. The status of this approximation was discussed in
detail in Ref.~\cite{ABG}.
For an open quantum dot with the large number of open
channels $\Nch\gg 1$ the interaction term can be treated within mean
field approximation, and the Hamiltonian $\hat {\cal H}_{\rm d}$ in
\req{Hd} can be further simplified:
\be
\label{2}
{\cal H}_{\rm d}= \bm{\psi}^\dag
\left[ \hat H + \sum_i\hat V_i\varphi_i(t) +\hat 1 eV_{\rm d}(t) \right] \bm{\psi},
\quad e V_{\rm d}(t)=\sum_i Z_i\varphi_i(t)+2E_{\rm c}\langle
N\rangle.
\ee
Here we introduced
the electric potential $V_{\rm d}(t)$ linear in the
quantum mechanical average $\langle N\rangle$ of the electron number $N$
in the dot.
Corrections to this mean field treatment were calculated
in Refs.~\cite{BA,GZ,Bnewint},

To determine the electric potential $V_{\rm d}(t)$ in the
dot, we have to define the quantum mechanical average $\langle N \rangle$
of the electron number $N$ in the dot.
In each particular moment of time the electron number $\langle N \rangle$
is not constant and its time evolution is described by the discontinuity equation
$e\dot N(t)=I_{\rm r}(t)+I_{\rm l}(t)$.
We estimate $V_{\rm d}(t)$ to the lowest order in $1/N_{\rm ch}\ll 1$,
and use $\Nl G_0$ ($\Nr G_0$) for the conductance of the left
(right) contact, $G_0=e^2/\pi$ is the quantum conductance.
Then, $\langle N(t)\rangle$ satisfies the
following equation
\be
\frac{d \langle N (t) \rangle }{dt}
= -  \gesc \langle N(t) \rangle
+ \frac{eN_{\rm l}}{\pi}\left(V_{\rm l} - V_{\rm d}(t) \right)
+ \frac{eN_{\rm r}}{\pi}\left(V_{\rm r} - V_{\rm d}(t) \right)
.
\label{110a}
\ee
The first term in \req{110a} is a diffusion term, describing the electron escape
from the dot with rate $\gesc$, where $\gesc^{-1}$ is the mean time
for an electron to escape the dot through one of the leads; below we
define $\gesc$ in terms of microscopic parameters of the system.
The last two terms in \req{110a} represent electron flux from the dot
due to the voltage difference
$V_{\rm l,(r)}-V_{\rm d}(t)$ across the contact of the left (right)
reservoir
and the dot. A discussion of the charge dynamics in quantum dots
can be also found in~\cite{PB05}.

Combining \req{110a} with the expression for $V_{\rm d}(t)$ from \req{2},
we obtain
\be
V_{\rm d}(t) = \frac{2 E_{\rm c}\Nch }{2 E_{\rm c}\Nch + \pi\gesc}
\frac{\Nl V_{\rm l}+\Nr V_{\rm r}} {\Nch}
+
\frac{\gesc + \partial_t}
{\gesc + 2 E_{\rm c} \Nch /\pi + \partial_t }\sum Z_i\varphi_i(t).
\label{elnuetral}
\ee
The characteristic energy scale governing the dynamics of the charge
is $E_{\rm c}\Nch /2\pi\propto G_0\Nch /C_{\rm d}$, $C_{\rm d}$
is the dot capacitance. Usually,
this scale is of the order of the Thouless energy $E_T$ and
significantly exceeds electron escape rate $\gesc$. Therefore, we
consider the limit, when both $\gesc$ and the frequency of the
external field $\omega$ are much smaller than $E_{\rm c}$, and use
the following equation for the electrostatic potential of the dot
\be
V_{\rm d}(t)\equiv V_{\rm d} = \frac{N_{\rm l} V_{\rm l}+N_{\rm r} V_{\rm r}}{\Nch}.
\label{Vd}
\ee
We conclude that the time-dependent perturbation \req{Hd}
can be chosen traceless,  $Z_i=0$, and the electric
potential in the middle dot is determined by the potentials of the
left and right reservoirs.


\subsection{Electric Current}
\label{sec:2.2}

The current through the dot is given in terms of the scattering
matrices $\hat {\cal S}(t,t')$ by the following expression
\be
\label{10}
\langle I \rangle =
e \int_0^{\To}  \frac{dt}{\To}
\int\! dt_1dt_2
\tr
\left\{
\hat\Lambda\left[
\hat {\cal S} (t,t_1)\hat f(t_1-t_2)
\hat {\cal S}^{\dagger}(t_2,t)
-\hat f (+0)\right]
\right\}.
\ee
The derivation of \req{10} can be found in
\cite{BTP94,VAA,PBSmat}, see also \ref{app:A}.
Here $\langle I \rangle$ stands for the quantum
mechanical and thermodynamic averages of the current operator (no ensemble
averaging!) and $\hat f(t)$ represents the
electron distribution function in the leads in time representation.
We consider the case when electrons in the leads are in thermal
equilibrium at temperature $T$, but the different voltages $V_{\rm l}$ and
$V_{\rm r}$ are applied to the left and right electron reservoirs.
Then, the matrix $\hat f(t)$ is
diagonal  $f_{\alpha\alpha}(t)=f_{\rm l(r)}(t)$,
if channel $\alpha$ belongs to the left (right) lead.
The function $f_{\rm l(r)}(t)$ is the Fourier transform
of the Fermi--Dirac distribution function:
\be
\label{12}
f_{\rm l(r)} (\tau)=e^{ieV_{\rm l(r)}\tau} f(\tau); \quad
f(\tau) =
 \int\limits_{-\infty}^{+\infty}
\frac{d\omega}{2\pi}\ \  e^{i\omega \tau}
\left\{ \frac{1}{e^{\omega/T}+1}-\frac{1}{2}\right\}=
\frac{i T }{2\sinh\pi T \tau}.
\ee
Here the traceless diagonal matrix $\hat\Lambda$ is introduced
\be
\label{11}
\Lambda_{\alpha\beta}=\delta_{\alpha\beta}\cases{\displaystyle
+\frac{N_{\rm r}}{N_{\rm ch}}, & if $1\leq \alpha \leq N_{\rm l}$;\cr
\displaystyle
-\frac{N_{\rm l}}{N_{\rm ch}}, & if $N_{\rm l}< \alpha \leq
N_{\rm ch}$,\cr}
\ee
and the scattering matrix $\hat {\cal S}(t,t')$
\be
\label{Sdef}
{\cal S}_{\alpha\beta}(t,t')=e^{ieV_{\rm d}(t-t')}
\left[\delta_{\alpha\beta}\delta(t-t')-2\pi i\nu
W^\dag_{\alpha n} \R_{nm}(t,t') W_{m\beta}\right],
\ee
is defined in terms of the Green  function $\R_{nm}(t,t')$
that satisfies the following equation
\be
\label{14}
\left(i\frac{\partial}{\partial t}-{\hat H} -\sum_i\hat V_i\varphi_i (t)+
i\pi\nu \hat{W}\hat{W}^\dag \right)\hat G^{(R)}(t,t')=
\delta(t-t'),
\ee
where the matrices $\hat{H}$, $\hat V_i$ and $\hat{W}$ were
introduced earlier, see Eqs.~(\ref{7}) and (\ref{2}).
The diagonal component $eV_{\rm d}$
of the electron Hamiltonian in the dot
is removed from the expression for the electron Green  function
$\R_{nm}(t,t')$ by
the  gauge transformation, represented by the exponential factor
in \req{Sdef}.

To the linear order in voltage across the dot $V=V_{\rm l}-V_{\rm
r}$, the dc electric current has the form
\be
\langle I\rangle =\Iph + gV.
\label{I1}
\ee
The first term represents the photovoltaic current, which flows
through the dot even at zero bias. The second term is linear in
voltage $V$ with factor $g$ being the dc conductance of the dot in the
presence of time-dependent perturbations $\hat V_i$. The linear in
$V$ contribution to the current in general may come from two
sources: i) the non-equilibrium distribution of electrons in the
leads and ii) change in the photovoltaic current $\Iph$ due to change in
the configuration of the electron wave functions when the bias is applied.
Due to the electro-neutrality condition \req{elnuetral}, the
voltages $V_{\rm l}$,  $V_{\rm d}$ and $V_{\rm d}$ enter only as
exponential factors to the expression for the electric current
\req{10}, and do not actually affect the structure of electron
wave functions in the dot. Therefore, only the non-equilibrium
current contributes to the linear in $V$ term.

The dc conductance $g$ of the dot can be represented in the form
\be
\label{17}
g   = g_{\rm cl}+
G_0
\int\limits_0^{\To}\!\frac{dt}{\To}\!\! \int\limits_{-\infty}^{+\infty}
\!\!dt_1dt_2
F(t_1-t_2)
\tr
\Big{\{}
\hat {\cal S}(t,t_1)\hat \Lambda {\cal S}^\dagger(t_2,t)\hat \Lambda
\Big{\}}, \quad
g_{\rm cl}= G_0 \frac{N_{\rm l}N_{\rm r}}{N_{\rm ch}}.
\ee
Here $g_{\rm cl}$ is the classical conductance of the dot,
$\To$ is the observation time,
$G_0=e^2/\pi\hbar$ is the quantum conductance for doubly degenerate electrons
in spin states and  $F(x)$
is the Fourier transform of the derivative of electron
distribution function:
\be
\label{18}
F(t)= \frac{\pi T t}{\sinh\pi T t}.
\ee

The expression for the photovoltaic current $\Iph$ can be obtained
from \req{10} by taking $V_{\rm l}=V_{\rm r}$.
Using the Wigner transform for the scattering matrix
\begin{equation}
\label{12b} \hat {\cal S}(t,t')=\int \hat {\cal S}_{\frac{t+t'}{2}}(\vare)e^{i\vare
(t-t')}\frac{d\vare}{2\pi},
\end{equation}
we write
\be
\label{12a}
\Iph = e\int\limits_0^{\To}\frac{dt}{\To}
\int d\tau\int\frac{d\vare}{2\pi} e^{i\vare \tau} f(\tau)
 \tr \left\{ \hat \Lambda \hat {\cal
S}^{}_{\frac{t}{2}+\frac{\tau}{4}}\left(\vare\right) \hat {\cal
S}^{\dagger}_{\frac{t}{2}-\frac{\tau}{4}}\left(\vare\right)\right\}.
\ee
For slow perturbations $\varphi_i$ with frequencies $\omega_i$
smaller than temperature $T$ or
the inverse eigenvalues of the time delay matrix~\cite{timedelay}
\be
\hat{\cal R}_{\vare}(\vare,t)=
[\partial_\vare\hat {\cal S}_t(\vare)]\hat{\cal S}^\dagger_t(\vare),
\label{timedelay}
\ee
we can expand the scattering matrices in \req{12a}
in $\tau$ and obtain
\be
\label{12c}
\Iph = e\int\limits_0^{\To}\frac{dt}{\To}
\int\frac{d\vare}{2\pi}\frac{1}{\cosh^2\vare/2T}
\tr \left\{ \hat \Lambda \left(
\frac{\partial {\cal S}^{}_t\left(\vare\right)}{\partial t} {\cal
S}^{\dagger}_t\left(\vare\right)- {\cal S}^{}_t\left(\vare\right)
\frac{\partial {\cal S}^{\dagger}_t\left(\vare\right)}{\partial t}
\right) \right\}.
\ee
Here, the scattering matrix in the Wigner representation is a
function of the perturbation itself and its time derivatives:
$\hat {\cal S}_t(\vare)=\hat {\cal
S}(\vare,\varphi_i(t),\dot\varphi_i(t),\dots)$ because the Green
function $\hat G^{(R)}(\vare,t)$ is a solution of the equation:
\begin{eqnarray}
\fl
\vare \hat G(\vare,t) & - &
\frac{1}{2}\left\{ \hat H-i\pi \nu \hat W\hat W^\dagger;
\hat G^{(R)}(\vare,t)\right\}
\nonumber
\\
\fl & +&
\sum_{k=0}^{\infty} \sum_i \frac{1}{2(2i)^k k!}\frac{d^k\varphi_i(t)}{dt^k}
\left( \hat V_i
\frac{\partial^k \hat G(\vare,t)}{\partial \vare^k}
+(-1)^k \frac{\partial^k\hat G(\vare,t)}{\partial \vare^k }\hat V_i\right) =
1,
\label{Wigner}
\end{eqnarray}
with $\{\hat A;\hat B\}=\hat A\hat B+\hat B\hat A$.
In the adiabatic
approximation the derivatives $d^k\varphi_i(t)/dt^k$ can be neglected and
the scattering matrix is determined by parameters $\varphi_i(t)$,
$\hat {\cal S}_t(\vare)=\hat {\cal
S}(\vare,\varphi_i(t))$:
\be
\label{12c1}
\Iph =\frac{ e  }{T_{\rm p}} \oint d\varphi_i
\int \frac{\tr\left\{ \hat \Lambda \ {\rm Im }
\hat {\cal R}_i(\vare,\varphi)
\right\}}{\cosh^2\vare/2T}\frac{d\vare}{2\pi},
\quad
\hat {\cal R}_i(\vare,\varphi)=
\frac{\partial \hat {\cal S}(\vare,\bf{\varphi})}{\partial \varphi_i}
\hat {\cal S}^\dagger(\vare,\bf{\varphi}).
\ee
The integral in \req{12c1} runs over the loop in the parameter space $\varphi_i$, and
$T_{\rm p}$ is the time for a system to complete this loop.
Particularly, for the perturbation characterized by two parameters
\be
\varphi_1(t)=X_1 \cos(\omega t),\ \varphi_2(t)=X_2 \cos(\omega t + \phi)
\label{eq:X}
\ee
the photovoltaic current is given by~\cite{Brouwer98}
\be
\Iph=\frac{e\omega}{2\pi^2}\int_A d\varphi_1 d\varphi_2{\rm  Im}\tr\left\{
\hat\Lambda \frac{\partial \hat{\cal S}}{\partial \varphi_1}
\frac{\partial \hat{\cal S}^\dagger}{\partial \varphi_2}
\right\},
\label{varAD}
\ee
where the integral runs over the inner part of the ellipse,
defined  by \req{eq:X}.
At finite frequencies, but still $\omega_i\ll T$, \req{12c} is still
applicable and can be rewritten in the form similar to \req{12c1},
if the parameter space $\varphi_i$ is extended to the phase space,
containing time derivatives of $\varphi_i(t)$ as well.

With the help of the equations of motion \req{14},
the expression for the photovoltaic
current can be rewritten in terms of  the Green functions
$\hat G^{R,A} (t,t')$:
\be
\label{19H}
\displaystyle
\Iph
= 2e i \pi \nu  \int\limits_0^{\To} \frac{dt}{\To}
\int\!\!\int\!\!dt_1dt_2 F^{\rm ph}_i(t_1,t_2)
\sum_i\tr  \left\{
\hat W^\dagger \hat \R (t,t_1)  \hat V_i
\hat \A(t_2,t)\hat W\hat \Lambda\right\},
\ee
which is more convenient in some calculations.
Here the function
\be
F^{\rm ph}_i (t_1-t_2)= \left[\varphi_i(t_1)-\varphi_i(t_2) \right]
f(t_1-t_2),
\label{Fph}
\ee
takes into account the probability of electron transitions
due to the perturbation $\hat V\varphi_i(t)$ for the
equilibrium electron distribution in the dot $f(t)$. We notice,
that taking higher order terms in $\hat V_i$ in $\hat{G}^{R,A}$
results in a new electron distribution function in the dot:
\be
\hat V_i F^{\rm ph}_i (t-t')\to
\hat \R (t,t_1)  \hat V_i F^{\rm ph}_i (t_1-t_2)
\hat \A(t_2,t').
\label{Feff}
\ee
The effective electron distribution function has a shape different
from the Fermi distribution function, see Sec.~\ref{sec:5}.

We note that due to ${\rm tr}\hat\Lambda=0$
the expression for the conductance cannot be represented in terms
of modified distribution function. As a result, see~\cite{VAcf},
conductance fluctuations are characterized by the electron
temperature in the reservoirs rather than by the electron temperature in the
dot. This statement was further investigated in~\cite{YKK}, where
the effect of time-dependent perturbation on two possible definitions
of the conductance was studied. It was shown, that the
Landauer conductance, defined as the linear response to the bias between
the reservoirs and given by \req{17}, is indeed characterized by the
electron distribution function in the leads. In other geometries
one can measure the linear
response of electric current to the internal perturbation of
the mesoscopic system by the dc electric field. Such response,
called the Kubo conductance, is sensitive to the actual distribution function
of electrons in the mesoscopic system.


\subsection{Current Noise}
\label{sec:2.3}

The current correlation function $S$ represents
fluctuations of the charge $Q=\int_0^{\To} I(t) dt$
transported through the dot over the observation time interval
$\To$
\be
\label{4.2}
S=\frac{\langle Q^2 \rangle-\langle Q \rangle^2}{\To}
=\int_0^{\To} \left(\langle
I(t)I(t')\rangle- \langle I(t)\rangle \langle
I(t')\rangle\right) \frac{dtdt'}{\To} .
\ee
Expression for $S$ in terms of the scattering matrices $\hat {\cal
S}$ can be derived in a similar way to the derivation of \req{10}
for current, and is outlined in \ref{app:B}.
For arbitrary distribution function
$f_{\alpha\beta}(t)=\delta_{\alpha\beta} f_\alpha(t)$
in the leads the current correlation function $S$ has the
form ($\delta_{t,t'}=\delta(t-t')$):
\begin{eqnarray}
\label{4.3}
\fl
S & = & \int_0^{\To} dtdt' \int dt_1dt_2
dt'_1dt'_2
 \tr\left\{\left( \hat {\cal S}^\dagger (t_2,t) \hat
\Lambda \hat {\cal S} (t,t_1')-\hat \Lambda
\delta_{t_2,t}\delta_{t,t_1'} \right) \hat f (t_1'-t_2')\right.
\nonumber
\\
\fl
& \times & \left.\left( \hat {\cal S}^\dagger (t'_2,t') \hat \Lambda
\hat {\cal S} (t',t_1)-\hat \Lambda
\delta_{t_2',t'}\delta_{t',t_1}
\right) (\hat 1 \delta_{t_1,t_2}-\hat f (t_1-t_2))\right\}.
\label{Svardef}
\end{eqnarray}

Below we consider the case of zero bias across the dot,
so that $f_\alpha(t)\equiv f(t)$, see \req{12}. We also assume
that the temperature of the system $T$ is finite and $T\To\gg 1$.
(The limit $T=0$ has some interesting properties and was discussed in
Refs.~\cite{Lpump,KApump}).
Then, $S$ can be  divided  into two parts:
\be
\label{4.4}
S = S_{{\rm NJ}} + S_{\rm P}.
\ee
Here, the second term  $S_{\rm P}$ is chosen in such a way, that in
the absence of time dependent perturbations this term vanishes
[$S_{\rm P} = 0$, see \req{4.5} below], and only the first term remains. The first
term describes the current noise due to thermal fluctuations
of electrons in the leads at temperature $T$ and is known as
the Nyquist-Johnson noise~\cite{Nn,Jn}.

The Nyquist-Johnson component of the noise can be written as
\be
\label{4.6}
S_{{\rm NJ}} = 2 g_{\rm cl} T -
\int\limits_0^{\To} \frac{dtdt'}{\To} \int
dt_1dt_2  f(t_1-t')\tilde f(t'-t_2)
 \tr\left\{ \hat \Lambda \hat{\cal S}(t,t_1) \hat
\Lambda \hat{\cal S}^\dagger(t_2,t) \right\},
\ee
where $\tilde f(t)=\delta(t)-f(t)$. The first term in \req{4.6}
represents the noise of a classical resistor
with resistance $1/g_{\rm cl}$. The second term in \req{4.6}
describes the contribution to the current noise from the quantum mechanical
corrections to the conductivity, cf. \req{17}. In
the absence of time-dependent perturbations, the second term
represents the quantum correction to the conductance, so that
the noise correlator has the form
$S_{{\rm NJ}} = 2 g  T$, where $g$ is the sample-specific
conductance of the dot, see \req{17}.

The external field changes the conductance of the dot, see
Sec.~\ref{sec:4}. Consequently, we can expect that the
Nyquist-Johnson contribution to the current noise is also
modified due to the external field. In particular, the ensemble
average $S_{{\rm NJ}}$ and fluctuations of $S_{{\rm NJ}}$ with
respect to different dot realizations
are suppressed by
time-dependent perturbation.

The second term,  $S_{\rm P}$, in \req{4.4} represents the noise of
the photovoltaic current \req{12a} and has the following form  in terms
of the scattering matrix $\hat{\cal S}(t,t')$:
\begin{eqnarray}
\fl
\nonumber
S_{\rm P} & = & e^2 \int_0^{\To} \frac{dtdt'}{\To}
\int dt_1dt_2dt'_1dt'_2 f(t_1-t_2) \tilde f(t_1'-t_2')
\\
\fl
& \times & \tr\left\{ \hat{\cal S}(t_2',t)\hat \Lambda
\hat{\cal S}^\dagger(t,t_1) \hat{\cal S}(t_2,t') \hat \Lambda
\hat{\cal S}^\dagger(t',t_1')-\Lambda^2
\delta_{t_2',t}\delta_{t,t_1}\delta_{t_2,t'}\delta_{t',t_1'}
\right\}.
\label{4.5}
\end{eqnarray}
To discuss the noise of  the photovoltaic current
in more detail, we  consider
the adiabatic limit, when
the eigenvalues of the time-delay matrix~\req{timedelay}
are shorter than both $1/T$ and
$1/\omega_i$, ($\omega_i$ is the frequency of external perturbation $\hat V_i$).
The ensemble average value of
$S_{\rm P}$ for arbitrary strength and frequency of the
perturbations was investigated in \cite{PVB,MBnoise} and is briefly discussed
in the end of Sec.~\ref{sec:5}.

In the adiabatic limit only electrons close to the Fermi energy
contribute to the current. Thus, we can
neglect energy dependence of the scattering matrix $\hat{\cal S}(\vare,t)$
in the Wigner representation \req{12b} and
substitute $\hat{\cal S}(t,t')=\hat{\cal S}_t(\vare=0)\delta_{t,t'}$
into \req{4.5}:
\be
\label{4.8}
S_{\rm P}=e^2\int_0^{\To} \frac{dtdt'}{\To}\tr
\left\{ \hat \Lambda^2 - \hat {\cal S}^\dagger_t
(0)\hat \Lambda \hat {\cal S}^{}_t(0) \hat {\cal S}^\dagger_{t'} (0) \hat
\Lambda  \hat {\cal S}^{}_{t'} (0) \right\} f(t-t')\tilde f(t-t').
\ee
We
observe that the temporal correlations in the current survive on time scales
comparable with the observation time $\To$ at low temperatures $T\To\ll 1$.
In this case the full
counting statistics is non-trivial and higher moments of the
current should be investigated, see Refs.~\cite{Lpump,KApump} for more detail.
At finite temperature $T$, the temporal correlations of the
current are suppressed on time scale of the order of $1/T$, see
\req{12}, and the counting statistics of the current becomes Gaussian and
is  described by the average value of the current $\Iph$, \req{12a}
and its noise $S_{\rm P}$, \req{4.5}.

Within a bilinear response to the perturbation \req{eq:X}
we obtain the following expression for the noise
\be
S_{\rm P}  = e^2 F_{\rm n}(T,\omega)
\left({\cal K}_{11}X_1^2+{\cal K}_{22}X_2^2
 + 2\cos\phi\ {\cal K}_{12}X_1^{}X_2^{\vphantom{1000}}\right).
\label{bilinearNoise}
\ee
Here coefficients ${\cal K}_{ij}$ are given by
\be
{ \cal K}_{ij} =
\tr \left\{ [\hat \Lambda; \hat {\cal R}_i][\hat {\cal R}_j;\hat
\Lambda]
\right\}
\label{Kij}
\ee
and the function $F_{\rm n}(T,\omega)$ represents
the probability of the absorption or emission of a
perturbation quantum with energy $\omega$:
\begin{eqnarray}
\fl
F_{\rm n} &=& \int \frac{d\varepsilon}{4\pi}
\left[ f({\varepsilon+ \case{1}{2}\omega})
{\tilde f}({\varepsilon- \case{1}{2}\omega})
- 2 f({\varepsilon}) {\tilde f}({\varepsilon})
+ f({\varepsilon- \case{1}{2}\hbar\omega})
{\tilde f}({\varepsilon+ \case{1}{2}\hbar\omega})
\right]
\nonumber
\\
\fl
&=&
\frac{\omega}{2\pi}\left(
\coth\frac{\omega}{2T}-\frac{2T}{\omega}\right).
\label{prob}
\end{eqnarray}
At low temperatures $T\ll \omega$, but still $T \gg 1/\To$,
 $F_{\rm n}=\omega/2\pi$. As  $T$ increases,  $F_{\rm n}$ decreases
$F_{\rm n} = \omega^2/T$.

Above we discussed the current noise in the situation when the
bias across the dot is zero. When a finite bias is applied, the
noise acquires dependent on the bias contribution called shot noise.
It was shown~\cite{AAL} that the shot noise
originate only due to quantum corrections to electron transport,
while the classical contribution to the transport does not lead to
the shot noise. Therefore, one can expect that
a time-dependent perturbation suppresses shot noise along with
any other quantum interference characteristics of electron
transport. Another interesting effect of microwave radiation on the
shot noise of open quantum dots was found by Lamacraft in Ref.~\cite{lamacraftSN}.
This effect results in cusps of the noise power when the bias $eV$
is a multiple of microwave frequency $\omega$: $eV=n\omega$ with
integer $n$.


\section{Ensemble of Open Quantum Dots}
\label{sec:3}

The exact form of the Hamiltonian \req{2} for quantum dots depends
on many microscopic parameters of the system, such as the shape of
the dot, position of impurities and is usually too complicated for
analysis. However, for many purposes the interesting
question is what the statistical properties of transport coefficients through
a quantum dot, rather than the corresponding values for each
particular sample. To describe statistical properties of quantum
dots, a random matrix theory turns out to be a productive tool.
The random matrix description of quantum dots is based on the
assumption, that the Hamiltonian of the dot, \req{2}, is
determined by $M\times M$ matrices $\hat H$ and $\hat V_i$ with
$\hat H$ being a random realization of
a hermitian matrix from the Gaussian ensemble~\cite{mehta}.
The matrix elements $H_{nm}(\Phi)$ of matrices from this ensemble
in the presence of magnetic flux $\Phi$ through the dot
are described by  the following  correlators
\be
\overline{H_{nm}(\Phi_1) H^*_{n'm'}(\Phi_2)}=
\frac{M\mls^2}{\pi^2}\left[L(\Phi_1-\Phi_2)
\delta_{nn'}\delta_{mm'}
+
L(\Phi_1+\Phi_2)\delta_{mn'}\delta_{nm'}\right]
,
\label{4}
\ee
Here $\overline{(\dots)}$ stands for the ensemble  averaging, and
$\mls$ is the mean level spacing of eigenvalues of $\hat H$.
For small  $\Delta\Phi$ function $L(\Delta \Phi) $ can be
estimated as
$L(\Delta \Phi) =1 - \kappa  \left(\Delta\Phi/\Phi_{\rm q}\right)^2$,
where $\kappa$ is a non-universal, sample-specific constant
of the order of unity, and $\Phi_{\rm q}=c/e$ is the flux
quantum~\cite{EfetovMF,BeenakkerRMP}.
At $\Phi_{1,2}=0$, the matrix $\hat H(0)$ belongs to the orthogonal
ensemble. As $\Delta\Phi$ increases, $L(\Delta\Phi)$ vanishes,
and $\hat H(\Phi)$ becomes a matrix from the unitary ensemble
when the second tern in \req{4} is equal to zero.
The microscopic justification of the random matrix description
\req{4} can be found in \cite{Efetov82,Efetov83,Efetovbook} for disordered systems
and in~\cite{MK97,AASA,AOS} in ballistic chaotic systems.

Matrices $\hat V_i$ can also considered as Hermitian  random
matrices. Below we disregard the fluctuations of the matrices
$\hat V_i$, and assume that $\hat V_i$ are real symmetric matrices,
belonging to an orthogonal ensemble. In this
case, we characterize perturbations $\hat V_i$ by parameters
\be
\label{5}
{C}_{ij}=\frac{\pi }{M^2 \mls} \tr\hat V_i\hat V_j.
\ee
We remind that $\tr \hat V_i=0$, see \req{elnuetral}.
The parameters $C_{ij}$ have the meaning of the level
velocities which characterizing the evolution of  an energy level
$\vare_{n}({\bf \varphi})$ under
the external perturbation $\sum_i\hat{V}_i\varphi_i(t)$, \cite{SA,AS}:
\be
\label{6}
\frac{2\mls}{\pi} C_{ij}=
\overline{
\frac{\partial \vare_\nu}{\partial \varphi_i}
\frac{\partial \vare_\nu}{\partial \varphi_j}
}
-
\overline{\frac{\partial \vare_\nu}{\partial \varphi_i}}
\
\overline{\frac{\partial \vare_\nu}{\partial \varphi_j}}.
\ee
Parameters $C_{ij}$ are also related to the transition rates
of electrons under perturbation $\hat V_i$. Indeed, the transition rate
$\gamma_{11}$ due to perturbation $\hat V_1$  is determined by the Fermi
golden rule:
\be
\gamma_{11}=\sum_m 2\pi |V_{1;nm}|^2\delta(\vare_n-\vare_m\pm
\omega)\sim \frac{\overline{ |V_{1;nm}|^2  }}{\mls}\simeq  \frac{C_{11}}{\pi}.
\ee
The first equality sign follows from the Fermi golden rule, the
second sign represents an estimate of the characteristic value of
the matrix elements $\overline{|V_{1;nm}|^2}  $ and the density of
states $1/\mls$, the last equation is the definition of $C_{11}$, cf.
\req{5}. If perturbations induce uniform electric fields $E_i$ in a quantum dot with
typical length $L$, parameters  $C_{ij}$ can be estimated as
$C_{ij}\simeq e^2 E_iE_j L^2/E_{\rm Th}$, where $E_{\rm Th}\sim
M\mls$ is the Thouless energy.

Below we show that all statistical transport
characteristics of quantum dots in the presence of time-dependent perturbations
are functions of parameters $C_{ij}$. Thus, even
though $C_{ij}$ are free parameters, measurements of several
transport characteristics~\cite{HFPMDH,DCMH,VDCM} allow one to eliminate the
uncertainty of $C_{ij}$.

For calculations of different correlation functions of transport parameters
over the ensemble of random Hamiltonians $\hat H$, we
use a diagrammatic technique, similar to one developed for disordered metals,
see~\cite{AGD}. In this Section, we briefly discuss
the basic elements of this diagrammatic technique.

\begin{figure}
\begin{center}
\epsfxsize=8cm \epsfbox{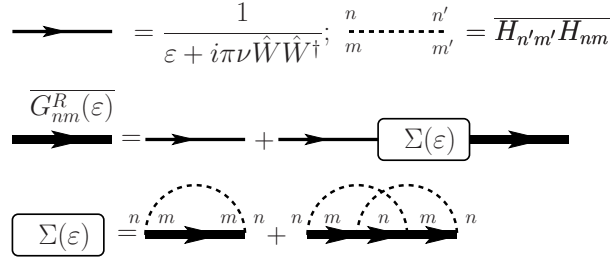}
\end{center}
\caption{\label{fig:1}
Diagrams for the ensemble averaged electron Green function in the dot.
The first line of the figure introduces the bare Green function
$[\vare+i\pi\nu\hat W\hat W^\dagger]^{-1}$ and correlation function
of the matrix elements of the Hamiltonian $\hat H$. The second
line represents the Dyson-type equation for the ensemble averaged
Green function $\overline{G_{nm}^{R}(\vare)}$. The third line
introduces the first two terms of the
self-energy $\Sigma(\vare)$, which is diagonal in index
of electron states in the dot.
The second term as well as all other terms which
contain intersections of dashed lines are
small in parameter $1/M$.
}
\end{figure}

First, we calculate
the ensemble averaged Green function $\hat G^{R,A}(\vare)$ in the absence
of time-dependent perturbations. The diagram equation in
Fig.~\ref{fig:1}
reduces to the following algebraic equation for the electron
self-energy $\Sigma(\vare)=(M\mls^2/\pi^2)\tr \hat
G^{R}(\vare)$:
\be
\Sigma(\vare)=\frac{M\mls^2}{\pi^2}\frac{1}{\vare-\Sigma(\vare)+i0}
-\Nch \frac{M\mls^3}{\pi^2}\frac{1}{\vare-\Sigma(\vare)+i0}
\frac{1}{\vare-\Sigma(\vare)+iM\mls/\pi}.
\label{selfcons}
\ee
Solving  \req{selfcons}, we find the ensemble average Green function
$\overline{\hat G^{R}(\vare)} =( \overline{\hat G^{A}(\vare)})^*$
for $\vare\ll M\mls$ in the form
\be
\label{15}
\overline{G_{nm}^{R}(\vare)}=- i\delta_{mn}\frac{\pi}{M\mls}
\cases{\displaystyle
 1+\frac{N_{\rm ch}+ i 2\pi\vare/\mls}{4M} , &
$N_{\rm ch}<n \leq M$; \cr
\displaystyle
\frac{1}{2}, & $1\leq n \leq N_{\rm ch}$.
}
\ee
In derivation of \reqs{selfcons} and \rref{15}, we used the following values for
factors $\Gm$ in \req{7}
\be
\label{16z}
\Gm= \sqrt{\frac{M\delta_1}{\pi^2\nu}}.
\ee
This choice of $\Gm$ corresponds to a dot connected to the
leads by reflectionless contacts, when the ensemble averaged
scattering matrix $\overline{{\cal S}_{\alpha\beta}}$
is zero and $\hat{\cal S}$ belongs to  circular ensemble, see~\cite{BeenakkerRMP}.

\begin{figure}
\begin{center}
\epsfxsize=12cm \epsfbox{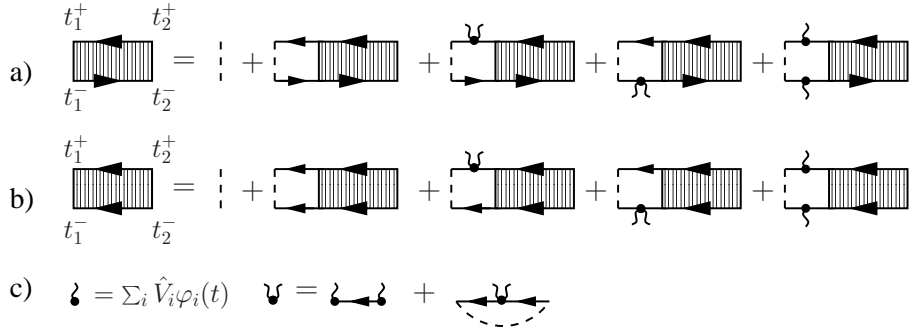}
\end{center}
\caption{\label{fig:2}
The Dyson-type equations for the diffuson, \req{diffdef}, and
the Cooperon, \req{coopdef},
are shown in the first and the second lines respectively.
The subscript ``${\rm amp}$'' in \reqs{diffdef} and \rref{coopdef}
emphasizes that the four Green functions at the terminals of the diffuson
and the Cooperon are omitted. All Green functions in these
diagrams and diagrams in figures below are ensemble average Green
functions, introduced in the second line of Fig.~\ref{fig:1}.
}
\end{figure}

We also introduce two other elements of the diagram technique
used in calculations of statistical properties of electron transport
in the presence of time-dependent perturbations.
One element  is called the diffuson $\D(t_1,t_2,\tau)$ and is
defined by
\begin{eqnarray}
\fl
\overline{{\big [}\R_{nm;\Phi_1}(t_1^+,t_2^+)
\A_{mn;\Phi_2}(t_2^-,t_1^-){\big ]}}_{\rm amp}
 & = & \frac{4M^2\mls^2}{\pi^2}\delta(t_1^+ + +t_2^- - t_2^+
 -t_1^-)
 \nonumber
 \\
\fl &&\times
\D_{\Phi_1-\Phi_2}
\left(\frac{t_1^++t_1^-}{2},\frac{t_2^++t_2^-}{2},t_1^+-t_2^+\right),
\label{diffdef}
\end{eqnarray}
see Fig.~\ref{fig:2}~a).
The other element is called the Cooperon
$\C(\tau_1,\tau_2,t)$ and  is defined as
\begin{eqnarray}
\fl
\overline{{\big [}\R_{nm;\Phi_1}(t_1^+,t_2^+)
\A_{nm;\Phi_2}(t_1^-,t_2^-){\big ]}}_{\rm amp}
 & = & \frac{4M^2\mls^2}{\pi^2}\delta(t_1^+ + +t_2^- - t_2^+
 -t_1^-)
 \nonumber
 \\
\fl &&\times
\C_{\Phi_1+\Phi_2}\left(t_1^+-t_1^-,t_2^+-t_2^-,\frac{t_1^++t_1^-}{2}\right),
\label{coopdef}
\end{eqnarray}
see Fig.~\ref{fig:2}~b).
These two elements represent the ensemble average product of the
advanced and retarded electron Green functions in the dot, divided
by the product $\overline{\R_{nn;\Phi_1}}
\
\overline{\R_{mm;\Phi_1}}
\
\overline{\A_{nn;\Phi_2}}
\
\overline{\A_{mm;\Phi_2}}$, so-called
the amputated average.
The diffuson and the Cooperon are given by the following
expressions:
\begin{eqnarray}
\fl
{\cal D}_{\Delta\Phi}(t_1,t_2,\tau) &  =  &\theta(t_1-t_2)
\exp\left( -\int\limits_{t_2}^{t_1}
\Gamma_{\Delta\Phi}(\tau,t)dt \right);
\label{diff}
\\
\fl
{\cal C}_{\Delta\Phi}(\tau_1,\tau_2,t) & = & \theta(\tau_1-\tau_2)\exp\left(
-\frac{1}{2}\int\limits_{\tau_2}^{\tau_1}\Gamma_{\Delta\Phi}(\tau,t)d\tau
\right).
\label{coop}
\end{eqnarray}
Here we use the notation
\begin{eqnarray}
\fl
&&
\Gamma_{\Delta\Phi}(\tau,t) = \esc+ \gamma(\Delta\Phi)
+\sum_{ij} \tilde\varphi_i(\tau,t) C_{ij}\tilde\varphi_j(\tau,t),
\label{Gamma}
\\
\fl
&& \esc =\frac{\Nch\mls}{2\pi};\ \ \ \
\gamma(\Delta\Phi)=\frac{2M\mls}{\pi}\left[1-L\right];
\ \ \ \
\tilde \varphi_i(\tau,t)=\varphi_i(t+\tau/2)-\varphi_i(t-\tau/2).
\label{miscGammas}
\end{eqnarray}
The first term in \req{Gamma} is the electron escape rate from the
dot
and $\gamma_{\Delta\Phi}$ is the electron dephasing rate due
to the difference in magnetic flux $\Delta\Phi$. The last term in
\req{Gamma} describes the effect of time-dependent field on the
correlation functions \reqs{diffdef} and \rref{coopdef} of
electron propagators.
Equations~\rref{15}, \rref{diff} and \rref{coop}
are the building blocks of the diagrams, which are studied below
for different correlation functions of transport characteristics
of open quantum dots.

\section{Effect of time-dependent perturbations on the conductance}
\label{sec:4}

\subsection{Weak localization correction}
\label{sec:4.1}

Weak localization correction to the conductance of a quantum dot
is given by the ensemble average of the second term in \req{17}.
For the unitary ensemble weak localization correction~\cite{LarkinWL} is small as
$g_{\rm cl}/\Nch^2\ll 1$ and is beyond the accuracy of our calculations.
In the orthogonal ensemble the weak localization
correction is $g_{\rm cl}/\Nch\sim 1$~\cite{BeenakkerRMP}. We define
the weak localization correction to the conductance as
the difference between the averaged values of the conductance
over orthogonal ($\Phi=0$) and unitary ($\Phi\gg \Phi_{\rm q}$) ensembles:
\be
\Delta g_{\rm wl}=\overline{(g)_{\Phi=0}}
-\overline{(g)_{\Phi\gg\Phi_{\rm q}}}.
\label{wldef}
\ee
In this subsection we
describe the effect of time-dependent field on the weak
localization correction \req{wldef}.

\begin{figure}
\begin{center}
\epsfxsize=5cm
\epsfbox{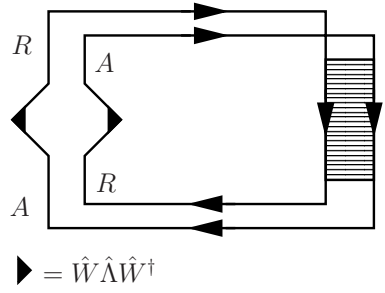}
\end{center}
\caption{\label{fig:WL}
The diagram for the calculation of the
weak localization correction to
the conductance.
}
\end{figure}

The weak localization correction is given by the diagram
in Fig.~\ref{fig:WL} and can be calculated from the following
expression~\cite{VAwl}:
\be \label{19}
\Delta g_{\rm wl}=
\Delta g^{(0)}_{\rm wl}
\int\limits_0^{2\pi/\o}\! \frac{ \omega d t}{\pi}
\int\limits_{0}^\infty\! \esc d\tau {\cal
C}(\tau,-\tau,t),
\quad
\label{190}
 \Delta g_{\rm wl}^{(0)}=-G_0\frac{\Nl\Nr}{\Nch^2} .
\ee
This equation  gives the universal description of the effect of the
time-dependent fields on the weak localization correction. Below we will
discuss different asymptotic regimes for the case when the
perturbation is described by only one harmonic function
$\varphi_1(t)=\cos\omega t$ and $C_{11}=C_{\rm l}$.

In the absence of the time dependent perturbation
$C_{\rm l} \equiv 0$, one obtains $\Delta g_{\rm wl}=
\Delta g^{(0)}_{\rm wl}$~\cite{BeenakkerRMP,BB96}.
For weak external field
$C_{\rm l} \ll \esc$ we find
\be
\label{24a}
\frac{ \Delta g_{\rm wl} } { \Delta g^{(0)}_{\rm wl} }=
1-\frac{\pi C_{\rm l}}{\esc}
\frac{\omega^2} {\omega^2+\esc^2},
\ee
where $\esc$ is defined in~\req{miscGammas}.
In this regime the correction is quadratic in the frequency of
slowly oscillating field, similarly to the result for bulk metal system at
$\omega$ smaller than the dephasing rate $1/\tau_\phi$. However, the
frequency dependence saturates at large frequency. It is different
from the result for bulk systems~\cite{AAKL}, where a characteristic spatial
scale shrinks as $\sqrt{D/\omega}$ with $D$ being the diffusion coefficient,
whereas in a quantum dot this scale is determined by the size of the dot
$L$. The random  matrix description breaks down at $\omega\sim D/L^2=E_{\rm
Th}$, $E_{\rm Th}$ is the Thouless energy.

In the opposite limit of strong external field $C_{\rm l} \gg \esc$
we consider separately the cases of fast, $\omega \gg\esc $, and slow,
$\omega \ll \esc$ oscillations. In the first case we
have
\be
\label{25}
\frac{ \Delta g_{\rm wl} } { \Delta g^{(0)}_{\rm wl}}=
\sqrt{\frac{\esc}{\pi C_{\rm l}}}.
\ee
The $1/\sqrt{C_{\rm l}}$ power dependence of the
quantum correction is similar to that for the bulk
system. Contrary to the bulk systems, the result does not depend on
the frequency $\omega$ for the reason mentioned above.
In the case of slow field $\omega \ll \esc$, but still $C_{\rm l}\omega^2 \gg \esc^3$
(strong field) the weak localization correction to the conductance
is
\be
\label{26}
\frac{ \Delta g_{\rm wl} } { \Delta
g^{(0)}_{\rm wl} }=\frac{\Gamma(1/6)}{\pi\Gamma(5/6)}
\sqrt[3]{\frac{\pi \esc^3}{9 C_{\rm l} \omega^2}}.
\ee
The power and frequency dependence of $\Delta g_{\rm wl} $
is again different from that in bulk disordered metals,
$\Gamma(x)$ is the $\Gamma-$function.

\subsection{Conductance fluctuations}
\label{sec:4.2}

\begin{figure}
\begin{center}
\epsfxsize=12cm
\epsfbox{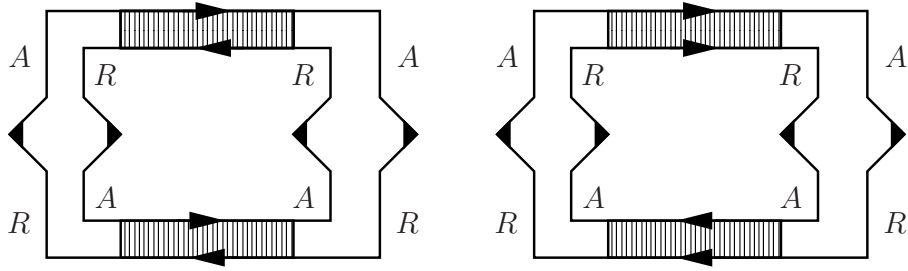}
\end{center}
\caption{\label{fig:CF}
The diagrams for the calculation of
the variance of  conductance fluctuations.
}
\end{figure}

Next we consider the fluctuations of the conductance $g$ over
ensemble of random Hamiltonians  $\hat H$. We notice that
the fluctuations in $g$ originate only from the second term in
\req{17}, since in the model of fully chaotic quantum dots with open channels
the classical conductance $g_{\rm cl}$ does not fluctuate. We have for
$\delta g=g-g_{\rm cl}$ the following correlation function, which
can be derived~\cite{VAcf} from the diagrams shown in Fig.~\ref{fig:CF}:
\be
\overline{\delta g_{\Phi_1} \delta g_{\Phi_2}}=
\frac{g_{\rm cl}^2}{\Nch^2}\esc^2\!\!
\int\limits_0^{2\pi/\omega}\!\! \frac{\omega^2dtdt'}{4\pi^2}
\int\limits_0^\infty\!\!  d\tau F^2(\tau)\int\limits_{\tau/2}^\infty
\!\!d\theta\left[
K^+(t,t',\tau,\theta)+K^-(t,t',\tau,\theta)
\right],
\label{CF}
\ee
where the functions $K^{\pm}(t,t',\tau,\theta)$ are given by
\begin{eqnarray}
\fl
K^+ (t,t',\tau,\theta) & = &
{\cal
D} \!\! \left(\frac{t+t'}{2},\frac{t+t'+\tau}{2}-\theta,t'-t\right)\!\!
{\cal
D} \!\! \left(\frac{t+t'}{2},\frac{t+t'-\tau}{2}-\theta,t-t'\right),
\label{Kplus}
\\
\fl
K^- (t,t',\tau,\theta) & =  & {\cal C}\left(t-t'+\theta-\frac{\tau}{2},t-t'-\theta+\frac{\tau}{2},
\frac{t+t'-\theta}{2}+\frac{\tau}{4}\right)
\nonumber
\\
\fl
&&\times {\cal C}\left(t'-t+\theta+\frac{\tau}{2},t'-t-\theta-\frac{\tau}{2},
\frac{t+t'-\theta}{2}-\frac{\tau}{4}\right).
\label{Kminus}
\end{eqnarray}

The two terms in \req{CF} have different properties with respect
to the magnetic flux $\Phi$ through the dot. Although at
$\Phi=0$ both terms survive, at finite magnetic field,
$|\Phi_{1,2}|\sim\Phi_{\rm q}$, only one of them remains:
for $\Phi_{2}\approx \Phi_{1}$ the second term vanishes, and for
$\Phi_{2}\approx -\Phi_{1}$ the first term vanishes. The
values of the conductance correlation function at $\Phi_{2}=\pm\Phi_1$
characterize the symmetry of the conductance with respect to
magnetic field inversion. If conductance is symmetric,
$\overline{\delta g_{\Phi} \delta g_{\Phi}}=
\overline{\delta g_{\Phi} \delta g_{-\Phi}}$.
This equation is indeed valid in the absence of time-dependent
perturbations~\cite{Onsagersymm,Bsymm}. As was shown in
Refs.~\cite{VAcf,WKcf},  time-dependent perturbations may
suppress the symmetry of the conductance with respect to
magnetic field inversion, see e.g. \req{ons2} below.

In the presence of a single harmonic perturbation at frequency
$\omega$ and with strength $C_{11}=C_{\rm l}$,
we can write the conductance correlation function in the form
\be
\label{24}
\overline{\delta g_{\Phi_1} \delta g_{\Phi_2}} =\frac{g_{\rm cl}^2}{\Nch^2}
\left[
\frac{\esc^2  }{\gamma_{-}^2}
Q^{+}\left(\frac{C_{\rm l}}{\gamma_{-}},\frac{T}{\gamma_{-}},
\frac{\omega}{\gamma_{-}}\right)
 +
\frac{\esc^2 }{\gamma_{+}^2}
Q^{-}\left(\frac{C_{\rm l}}{\gamma_{+}},\frac{T}{\gamma_{+}},
\frac{\omega}{\gamma_{+}}\right)\right] ,
\ee
where we used a shorthand $\gamma_\pm=\esc+\gamma(\Phi_1\pm\Phi_2)$ with
$\esc$  and $\gamma(\Delta\Phi)$ defined in \req{miscGammas}.
At $T=0$, $\Phi_{1,2}=0$, and in the absence of time-dependent perturbations
$Q^\pm=1$.
Below we discuss the properties of the functions $Q^{\pm}$ in various
regimes.

In the limit of high temperature, $T \gg \esc$, we obtain
\be
\label{27}
Q^\pm(x,y,z)\approx \frac{\pi^2}{3 y}\frac{1}{\sqrt{1+2x}}.
\ee
The equality between functions $Q^{\pm}$ means that the
conductance is symmetric with respect to magnetic field inversion.
However, in low temperature limit $T \ll \gamma_{\pm}$, we obtain for
strong perturbation $C_{\rm l}\gg \esc$
\be
\label{28}
Q^{+}(x,0,z) \approx  \frac{1}{2\sqrt{2x}},\quad
\label{28a}
Q^{-}(x,0,z) \approx  \frac{1}{2x}.
\ee

Equation~\rref{24} with $Q^{\pm}$ given by  \req{28} shows an
important signature of the effect of time-dependent perturbations
on the conductance --- the violation of the Onsager symmetry:
\be
\frac{\overline{\delta g_{\Phi} \delta g_{-\Phi}}}{
\overline{\delta g_{\Phi} \delta g_{\Phi}}} =
\sqrt{\frac{2\esc}{C_{\rm l}}},\quad \gamma(2\Phi)\gg \esc.
\label{ons2}
\ee
This breakdown of the Onsager relation is a simple manifestation of
lifting of the time reversal symmetry in the system with time
dependent Hamiltonian.

In the limit of low frequency $\omega\ll \esc$,
the conductance $g$ can be represented as the result of averaging
of the conductance $g(\{\varphi_i\})$ at stationary perturbation
$\varphi_i$ over one period $2\pi/\omega$:
\be
g=\int_0^{2\pi/\omega} g(\{\varphi_i(t)\}) \frac{\omega dt}{2\pi},
\label{gad}
\ee
where $g(\{\varphi_i\})$ can be calculated according to \req{17}
with the scattering matrix defined by \reqs{Sdef} and \rref{14}
at  fixed values $\varphi_i$. Because $g(\varphi)$
has magnetic field symmetry~\cite{Bsymm},
$g$ is also symmetric with respect to inversion of magnetic field.
Calculations of \req{CF} at $\omega\ll \esc$ give
\be
Q^{\pm}(x,y,0)= \int_0^{2\pi}\frac{d\xi d\zeta}{4\pi^2}\int_0^{\infty}
F^2(\xi/\gamma_{\mp})
\frac{\exp(-(1+4x\sin^2\xi/2\sin^2\zeta/2)\xi)}
{1+4x\sin^2\xi/2\sin^2\zeta/2} d\xi.
\ee
This expression in the limit of high temperature $T\gg \esc$
has the  asymptote
\be
\label{31a}
Q^{\pm}(x,y,0)=\frac{\pi}{3 y}K\left(-4x\right),
\ee
and at zero temperature $Q^\pm(x,0,0)$ is given by
\be
\label{31}
Q^{\pm}(x,0,0)=\frac{1}{\pi}\frac{E\left(-4x\right)+(1+4x)K(-4x)}{1+4x},
\ee
where $K(x)$ and $E(x)$ are the elliptic integrals of the first and
second kind respectively.

Suppression of the conductance fluctuations by slow field $\omega\ll \esc$,
see \req{31a} and \rref{31} is the consequence of
averaging of the stationary conductance $g(\varphi)$ over
different configurations of the full Hamiltonian along the closed contour
in parameter space, see Fig. 1 b).
Thus, the observed d.c. conductance $g$, \req{gad}, is already partially averaged
over ensemble of random Hamiltonians $\hat H$ and its fluctuations are reduced.
As the strength of the perturbation $C_{\rm l}$ increases, more
statistically independent configurations of Hamiltonian $\hat H$
contribute to the conductance $g$ and fluctuations of $g$ become
suppressed.

However, low-frequency  perturbations do not affect the weak
localization correction to the conductance $\Delta g_{\rm wl}$,
\req{wldef}, which is
defined as the difference between the averages over orthogonal
and unitary ensembles. Only
perturbations at frequencies $\omega\sim \esc$ could suppress
$\Delta g_{\rm wl}$, see e.g. \req{24a}, when
the conductance $g$, \req{17}, is no longer related to
the stationary conductance $g(\varphi)$. In this case
the suppression of both conductance fluctuations and
the weak localization correction to the conductance are qualitatively
similar and can be interpreted as dephasing.

\section{Photovoltaic current}
\label{sec:5}

\begin{figure}
\begin{center}
\epsfxsize=6cm
\epsfbox{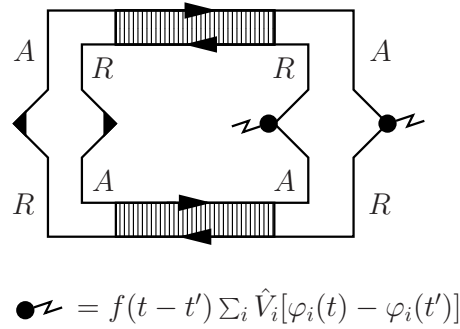}
\end{center}
\caption{\label{fig:AP}
The diagram representing the contribution to the variance of
the photovoltaic current, $\varIp$, at high temperature $T$.
}
\end{figure}

Photovoltaic current averaged over ensemble of
random Hamiltonian $\hat H$ is zero, because there is no specific
direction for the current to flow. However, for each particular
configuration of the quantum dot, a finite current can flow in
either direction. To characterize the value of this current,
one can find~\cite{VAA,PBSmat}
\be
\varIp=\frac{\omega^2 e^2}{4\pi^2}\frac{\Nl\Nr}{\Nch^2}
\int\limits_0^{2\pi/\omega}\frac{\omega^2 dtdt'}{\pi^2}
\int\limits_0^\infty\esc^2d\tau\int\limits_{\tau/2}^\infty
d\theta K^+(t,t',\tau,\theta) B(t-\theta,t'-\theta,\tau),
\label{varIph}
\ee
where $K^+$ is defined in \req{Kplus} and
\begin{eqnarray}
\fl
&& B(t,t',\tau) = \esc^2
  f^2(\tau)  \int\limits_0^{\infty}d\xi d\xi'
  {\cal D}(t,t-\xi,\tau){\cal D}(t',t'-\xi',\tau)
\left[\sum_{ij}\frac{C_{ij}}{\esc}\tilde\varphi_i(\tau,t)
\tilde\varphi_j(\tau,t')\right.\nonumber
  \\
\fl &&
+\left.
2\left(
\sum_{ij}\frac{C_{ij}}{\esc}\tilde\varphi_i(\tau,t-\xi)
\tilde\varphi_j(\tau,t-\xi)
\right)
\left(
\sum_{ij}\frac{C_{ij}}{\esc}\tilde\varphi_i(\tau,t'-\xi')
\tilde\varphi_j(\tau,t'-\xi')
\right)
\right]
\label{Bdef}
\end{eqnarray}
with $\tilde \varphi_i(\tau,t)$ introduced in \req{miscGammas}.
In Fig.~\ref{fig:AP} we
present only the diagram which survives at high temperatures
and the full set of diagrams contributing to \req{varIph} can be found
in Refs.~\cite{VAA,PBSmat},
We emphasize that the photovoltaic current has no symmetry with
respect to inversion of magnetic field,
$\overline{\Iph(\Phi)\Iph(-\Phi)}=0$, this statement in the
diagrammatic language means that there is no counterpart of the
diagram in Fig.~\ref{fig:AP}, that contains the Cooperons;
cf.~Fig.~\ref{fig:CF}.

Function $B(t,t',\tau)$, \req{Bdef}, is related to the
electron distribution function in the dot. Particularly,
at high temperature $T\gg T_{\rm h}$ for
harmonic perturbations at frequency $\omega$
\be
B(t,t',\tau)=
\sum_{ij}C_{ij}\dot\varphi_i\left(\frac{t+t'}{2}\right)
\dot\varphi_j\left(\frac{t+t'}{2}\right)
\left(\frac{ T \sin(\omega\tau/2)}{\omega^2\sinh  \pi T \tau}\right)^2,
\label{BhT}
\ee
corresponds to a square of the distribution function
$F^{\rm ph}_i$, \req{Fph}.
Here we introduced temperature scale $T_{\rm
h}$ according to
\be
T_{\rm h}=\omega{\rm max}\
\left\{\sqrt{\frac{C_{ij}}{\gamma_{\rm esc}}}\right\}.
\label{Th}
\ee
For a single perturbation at high frequency
$\omega\gg \esc$ with strength $C_{\rm l}$ and at low temperature
$T\ll T_{\rm h}$, we can estimate the integrals
over $\xi$ and $\xi'$ in \req{Bdef} as
\be
\label{43}
\int_0^\infty \D(t,t-\xi,2\tau)d\xi\approx
\frac{\esc}{\esc+2C_{\rm l}\sin^2\omega\tau},
\ee
and the function $B(t,t',\tau)$ acquires extra factors~\req{43}, which
vanish at $\tau\gg 1/T_{\rm h}$. Narrowing of the distribution
functions in time representation means the broadening of electron
distribution in the energy space. Indeed,
energy of an electron in the dot changes due to the external
field. Such changes
result  in the redistribution of the electrons in the energy
space and the new distribution function becomes wider than that of
electrons in the leads at temperature $T$.

The new width of the electron distribution
function can be estimated from the following argument.
The transitions occur with rate $C_{\rm l}$, and electron stays in the dot
for $1/\esc$ time, so that it experiences of the order
$C_{\rm l}/\esc$ transitions.
After each transition electron energy changes by $\pm\omega$, and
assuming that its motion in the energy space can be described by a
random walk, we find that on average electron energy changes by
$T_{\rm h}=\omega\sqrt{C_{\rm l}/\esc}$, which is consistent with the
estimate \req{43}.
The above estimate of $T_{\rm h}$ has a meaning only for strong fields,
$C\gg \gamma_{\rm esc}$, so that the diffusion picture in energy space is valid.
Otherwise, electrons experience only a few transitions with
energy change $\omega$.
Note that function in \req{43} is periodic in $\tau$ with
period $\sim 1/\omega$, i.e. at longer times the electron
diffusion is no longer described by random walk and some  structure
in the distribution function appears~\cite{shytov}.

For weak harmonic perturbation $C_{ij}\ll \esc$,
we expand $K^+$ to first order in $C_{ij}$ and neglect the second term in \req{Bdef},
since $T_{\rm h}\ll T$. As the result we obtain
for the harmonic perturbation, characterized by two
functions $\varphi_1(t)=\cos\omega t$ and $\varphi_2(t)=\cos(\omega t+\phi)$:
\be
\label{27ad}
\displaystyle \varIp = e^2\omega^2 \frac{\Nl\Nr}{\Nch^2}
\int\limits_0^{\infty}\!\!\esc  d\theta e^{-2\esc \theta}
\int\limits_{-\theta}^{+\theta} d\tau B(\tau)
\frac{C_{\rm l}^2 (2\omega \theta
-\sin 2\omega \theta ) + C_{\rm c}^2 \sin 2\omega \theta }{\omega},
\ee
where the linear, $C_{\rm l}$, and circular, $C_{\rm c}$,
perturbation amplitudes were introduced according to
\be
\label{29}
C_{\rm l} = C_{11}+2C_{12}\cos \phi+C_{22},\quad
C_{\rm c} =  2 \sin\phi \sqrt{C_{11}C_{22}-C_{12}^2}.
\ee
Note that the separation of the variance of the current into circular and linear
contributions corresponds to the classification of the current
components introduced in Ref.~\cite{FK}, where current through
a microjunction in microwave field with linear and circular
polarizations
was studied.
In the case of temperature $T$ larger than the escape rate,
$T\gg \gamma_{\rm esc}$, we have
\be
\label{30}
 \varIp = \frac{\pi}{12}\frac{e^2\omega^2}{4\pi^2} \frac{\Nl\Nr}{\Nch^2}
 \frac{\esc}{T}
\frac{1}{\esc^2+\omega^2} \left(
\frac{\omega^2}{\esc^2}C^2_{\rm l}
 + C_{\rm c}^2
\right).
\ee
The second term of Eq.(\ref{30}) survives the limit $\omega\to 0$,
thus reproducing the known result for adiabatic pumping
\cite{Brouwer98,SAA}. On the other hand, this term vanishes at high
frequency. The $C_{\rm l}^2$ term is quadratic in frequency at small
frequency and tends to a constant at large frequency.

The linear pumping amplitude $C_{\rm l}$ in the case of two pumps
has the form of Eq.(\ref{29}), which implies that the amplitude
$C_{\rm l}$ is
just a vector sum of different pumps in the parameter space. On the
other hand the circular amplitude is determined by
``uncorrelated'' components of matrices $\hat V_{1,2}$ and
vanishes if $\hat V_2\propto \hat V_1$.

To describe the variance of the photovoltaic current,
we first consider the adiabatic limit, when parameters $C_{ij}$
have a special form $C_{11}=C_{22}=C$ and
$C_{12}=0$. In this case the expression for the variance
can be written in the compact form for $T\gg\esc$
\be
\label{25ht}
\varIp = \frac{e^2\omega^2}{24\pi} \frac{\Nl\Nr}{\Nch^2}
\frac{\esc}{T}
\frac{2C+(\gamma_{\rm esc}- \sqrt{\gamma_{\rm esc}(\gamma_{\rm
esc}+4C)})} {\sqrt{\gamma_{\rm esc}(\gamma_{\rm esc}+4C)}}.
\ee
As temperature drops down to $T\ll \gamma_{\rm esc}=N_{\rm
ch}\mls/2\pi$, the variance of $\Iph$ saturates to
\be
\label{25lt}
\varIp=\frac{e^2\omega^2}{\pi^2}\frac{\Nl\Nr}{\Nch^2}
\frac{C^2}{\sqrt{\gamma_{\rm esc}(\gamma_{\rm esc}+4C)^3}}.
\ee
The authors of Ref.~\cite{SAA} showed that at strong
perturbation the variance of
the photovoltaic current in the adiabatic limit
is proportional to the perimeter of the contour integral
in the parameter space Fig. 1 b); $\varIp\propto \sqrt{C}$. This perimeter
law is the consequence of the lack of
correlation between distant points of the contour in the
parameter space.  The total contribution
to the pumped current consists of uncorrelated contributions
of the loop and is proportional to the number of independent
contributions $\sqrt{C/\esc}$.
In the opposite case of weak perturbation $C\ll\esc$ the current
$\Iph$ is determined by \req{varAD}, see~\cite{Brouwer98} and
is proportional to $C_{\rm c}$ ($\varIp\propto C_{\rm c}^2$).
Equations~(\ref{25ht}) and (\ref{25lt}) are consistent with the above arguments for
power dependence of $\varIp$.

When only one perturbation $\varphi_i(t)$ with power $C_{\rm l}$
is applied, the photovoltaic current vanishes in the adiabatic limit.
In this case the photovoltaic current is quadratic in
frequency $\omega$ for  $\omega\ll \esc$. For weak pumping $\varIp$
is determined by \req{30} with $C_{\rm c} =0$ for arbitrary frequency
$\omega$. For strong pumping $C_{\rm l}\gg \gamma_{\rm esc} $,
but still low frequency limit
$\omega^2C_{\rm l}\ll \gamma_{\rm esc}^3$, we have
\be
\label{32}
\varIp = \frac{25}{288} \frac{e^2\omega^2}{4\pi^2} \frac{\Nl\Nr}{\Nch^2}
\frac{\omega^2}{\esc^2} \frac{\esc}{T}\left(\frac{C_{\rm
l}}{\esc}\right)^{3/2}.
\ee
In the limit of high frequencies, $T\gg \omega\gg \gamma_{\rm esc}$,
the variance of the photovoltaic current is given by
\be
\label{33}
\varIp = \frac{e^2\omega^2}{24\pi} \frac{\Nl\Nr}{\Nch^2} \frac{\esc}{T}
\frac{C_{\rm l}+\gamma_{\rm esc}- \sqrt{\gamma_{\rm esc}(\gamma_{\rm
esc}+2C_{\rm l})}} {\sqrt{\gamma_{\rm esc}(\gamma_{\rm esc}+2C_{\rm
l})}}.
\ee
In the limit of strong pumping this expression has the
$\sqrt{C_{\rm l}}$ asymptotic behavior.

The results for $\varIp$ at finite frequency have the following
interpretation. Based on \req{Wigner}, we can
represent the photovoltaic current in the form similar to \req{12c1},
where the contour of integration is considered in the phase space.
The phase space includes
both parameters $\varphi_i(t)$ and their time derivatives~\cite{VAA}, as follows from
\req{Wigner}.
At weak perturbation with a single parameter $\varphi_1(t)=\cos\omega t$,
the contour is an ellipse with
semi-axes  proportional to $\sqrt{C_{\rm l}}$ and $\omega \sqrt{C_{\rm l}}$;
then $\Iph\propto (e\omega) \omega C_{\rm l}$ and $\varIp$ is consistent
with \req{30} at $\omega\ll\esc$.
In the limit of  strong perturbation at low frequency
$\omega^2C_{\rm l}\ll\esc^3$, the
contour in phase plane is long along the $\varphi_1$ axis but narrow in
the $\dot \varphi_1$ direction. The variance of the photovoltaic
current is determined by a sum of
independent contributions from pairs of the contour along the
$\varphi$ axis, the number of these pairs can be estimated as
$\sqrt{C_{\rm l}/\esc}$. Each pair consists of two adjacent pieces of the contour
shifted with respect to each other along $\dot\varphi_1(t)$ axis
and contributes as $\omega \sqrt{C_{\rm l}}$ to
the total current. As a result, we obtain $\varIp\propto (e\omega)^2\omega^2C_{\rm
l}^{3/2}$. Finally, if the amplitude of the field $C$ or the frequency $\omega$
increases further, $\omega^2 C_{\rm l} \gg \gamma^3_{\rm
esc}$, the contour does not have adjacent parts and each part of the contour
gives an independent contribution. Since the number of this parts is
$\sqrt{C_{\rm l}/\esc}$, the variance of the photovoltaic current
is proportional to $(e\omega)^2 \sqrt{C}$, see Eq.~(\ref{33}).

As frequency $\omega$ and power increase further, the heating
effects become important. At $T_{\rm h}\gg T$, the variance of the
photovoltaic current can be roughly estimated if electron
temperature $T$ in the leads is replaced by $T_{\rm h}$.
Particularly, from \req{33} we obtain the characteristic scale for
$\varIp$:
\be
\label{45} \varIp  \sim \frac{e^2\omega^2}{4\pi^2} \frac{\Nl\Nr}{\Nch^2}
\frac{\esc}{\omega}.
\ee
A numerical analysis~\cite{VDCM} shows that in fact at $C_{\rm l}\gg\esc$, the
variance of the photovoltaic current on the perturbation power has
a very weak (log-like) dependence on $C_{\rm l}$,
with typical of value $\varIp$ consistent with the estimate
of \req{45}.

The heating effects manifest themselves in the noise of the
photovoltaic current as well~\cite{PVB}. In the limit of strong perturbation
$C_{\rm l}\gg \esc$ at high frequency $\omega\gg \esc$ the ensemble averaged value
of $S_{\rm P}$, \req{4.5}, is
\be
\label{4.xx}
\overline{S_{\rm P}} \propto g_{\rm cl} T_{\rm h}.
\ee
The noise of the photovoltaic current has a
form similar to the expression for the Nyquist--Johnson noise, see
\req{4.6}: the current noise correlation function is
determined by the conductance of the dot $g_{\rm cl}$,
and the effective electron temperature. Due to the heating
by a strong perturbation, the electron distribution function is
broadened and the new energy scale for the electron distribution
function is given by $T_{\rm h}$, see \req{4.xx}. Thus, the
noise of the photovoltaic current averaged over the ensemble
has a similar origin with the Nyquist--Johnson noise and is determined by
thermal fluctuations of electrons in the dot out of equilibrium.


\section{Conclusions}


In summary, we reviewed the random matrix description of
electron transport through an open quantum dot, subject to time-dependent
perturbations. We expressed the dc current through the dot in terms
of the scattering matrices, and considered such components of the
current as the photovoltaic current, independent from the bias
voltage,
and the linear in the bias current, characterized by the conductance. The scattering
matrices are calculated in terms of time-dependent Hamiltonian,
that belongs to a Gaussian ensemble of random matrices. We then
presented the diagram technique to perform ensemble averaging  and
applied this technique to calculate different statistical
properties of the electron transport through the dot.

The main results can be summarized as follows. The weak
localization correction to the conductance and conductance
fluctuations are both suppressed by time-dependent perturbation.
However, the suppression has different parametric dependence on
perturbation frequency. The photovoltaic current can be represented as a
sum of circular and linear terms. These term have different
frequency dependence: the circular term dominates at low
frequencies and represents the adiabatic charge pumping, while the
linear term dominates at high frequencies. The photovoltaic
current and its noise are determined by the actual width of the
electron distribution function in the dot, on the other hand, the
variance of the conductance fluctuations is determined by electron
temperature in the leads. These results are in qualitative agreement with
experiments, described in Refs.~\cite{HFPMDH,DCMH}.

We described calculations using the Hamiltonian approach
to the statistical
description of the electron transport, a detailed description of
the scattering matrix approach for time-dependent system
can be found in Ref.~\cite{PBSmat}, where the same results were
obtained.

In this review we considered the electron system neglecting the
interaction effects and assumed spin degeneracy. The effect of
electron-electron interaction can be disregarded only in the limit
of the large number of open channels, but as the number of open
channels decreases, the interaction effects become more
important~\cite{BA,GZ,Bnewint,Bnewint1}. The interplay of the
interaction and time-dependent perturbation was addressed in
~Refs.~\cite{MBPRB64,CB}.

In semiconductor quantum dots in absence of magnetic field
electron spin states are nearly degenerate. However, if magnetic
field is applied, the spin degeneracy is lifted and currents of
electrons with opposite spin orientations are not identical. In
this case a spin current can be generated by time-dependent
perturbation, similar to the photovoltaic charge
current~\cite{spincurrent}, this effect was studied experimentally
in Ref.~\cite{spincurrexp}.
Another modification of the system, considered in the present
review, is a quantum dot connected to superconducting leads and
was studied theoretically in~\cite{sc1,sc2}, the experimental
realization of such a system remains a challenging task.

\section*{Acknowledgements  }
I would like to thank I. Aleiner, V. Ambegaokar, P. Brouwer, L.
DiCarlo, C. Marcus and M. Polianski, with whom I had a pleasure of
working
on various projects related to the topic of this review.
Discussions with B. Altshuler, M. B\"uttiker, A. Clerk, V. Falko, and V.
Kravtsov and A. D. Stone are greatly appreciated.
This work was supported by the
W. M. Keck Foundation and by NSF Materials Theory grant DMR-0408638.


\appendix


\section{}
\label{app:A}

We denote the wave function of electrons in channel $\alpha$
by $\psi_{\alpha}(x,t)$ with $x<0$ for incoming electrons and
$x>0$ for outgoing electrons, see Fig.~1 a).
The
boundary $x=0$ is described by a superposition of the incoming and
outgoing  electron states and we denote it by $\psi_{\alpha}(0,t)$.
The wave function of electrons in the dot is denoted by $\psi_i(t)$.

We introduce the matrix Green  function
\be
\label{a1}
\hat {\cal G}_{\alpha\beta}(t,t',x,x') = \left(
\begin{array}{cc}
{\cal G}^{(R)}_{\alpha\beta}(t,t',x,x') &
{\cal G}^{(K)}_{\alpha\beta}(t,t',x,x')\\
0 & {\cal G}^{(A)}_{\alpha\beta}(t,t',x,x')
\end{array}
\right),
\ee
which is defined in terms of the retarded, advanced and Keldysh
components as
\begin{eqnarray}
\fl
{\cal G}^{(R)}_{\alpha\beta}(t,t',x,x') & = &
-i\Theta(t-t')\langle \{\psi_{\alpha}(x,t);
\psi^\dag_{\beta}(x',t')\}\rangle,
\nonumber
\\
\fl
{\cal G}^{(A)}_{\alpha\beta}(t,t',x,x') & = &
i\Theta(t'-t)\langle \{\psi_{\alpha}(x,t);
\psi^\dag_{\beta}(x',t')\}\rangle,
\nonumber
\\
\fl
{\cal G}^{(K)}_{\alpha\beta}(t,t',x,x') & = &
-i\langle [\psi_{\alpha}(x,t);
\psi^\dag_{\beta}(x',t')]\rangle,
\nonumber
\end{eqnarray}
where $[A;B]=AB-BA$ and $\{A;B\}=AB+BA$. The similar expressions can be written down for
$\hat {\cal G}_{i\alpha}(t,t',x')$ Green  function, with
$\psi_{\alpha}(x,t)$ replaced by $\psi_i(t)$.

For non-interacting electrons, moving towards the dot ($x;\ x'<0$),
the Green's function is:
\be
\label{a3}
{\cal G}_{\alpha\beta}(t,t',x,x')= \left(
\begin{array}{cc}
\R_{\alpha\beta}(t-t',x-x') & \K_{\alpha\beta}(t-t',x-x')\\
0 & \A_{\alpha\beta}(t-t',x-x')
\end{array}
\right),
\ee
where
\begin{eqnarray}
\label{a4}
\fl
\R_{\alpha\beta}(t,x)&=&i\Theta(t)\ \delta_{\alpha\beta} \delta
\left( v_F t-x \right),\\
\label{a5}
\fl
\A_{\alpha\beta}(t,x)&=&-i\Theta(-t)\ \delta_{\alpha\beta} \delta
\left( v_Ft-x\right),\\
\label{a6}
\fl
\K_{\alpha\beta}(\vare,x)&=&  \tilde f_\alpha(\vare)\left(
\R_{\alpha\beta}(\vare,x)-\A_{\alpha\beta}(\vare,x)
\right),
\end{eqnarray}
and $ f(\vare)$ is the distribution function of electrons in channel
$\alpha$. If incoming electrons are in equilibrium at temperature $T$,
\be
\tilde f_\alpha(\vare)
=\tanh\frac{\vare- eV_\alpha}{2T},
\ee
with $V_\alpha$ being the voltage applied to the reservoir connected to the dot
by channel $\alpha$.

The equations of motion for the Green functions
$\hat {\cal G}_{\alpha\beta}(t,t',x,x')$ and
$\hat {\cal G}_{j\alpha}(t,t',x')$ are
\begin{eqnarray}
\label{a7}
\fl i\left[\frac{\partial}{\partial t}-v_F\frac{\partial}{\partial
x}\right]
\hat {\cal G}_{\alpha\beta}(t,t',x,x')
 & = & \delta(x)W_{\alpha i}\hat {\cal G}_{i\beta}(t,t',x')
+\delta(t-t')\delta(x-x')\hat 1,
\\
\label{a8}
\fl \ \ \ \
\left[i\frac{\partial}{\partial t}-H_{ij}(t)\right]\hat {\cal
G}_{j\alpha}(t,t',x')
& = &  W^\dag_{i\beta}\hat {\cal G}_{\beta\alpha}(t,t',0,x').
\end{eqnarray}
Due to causality principle, $\A_{\alpha\beta}(t,t',0,x')\equiv 0$
for $x'<0$. This observation
significantly simplifies further calculations. Indeed, we can
represent the Keldysh component of the Green's function in the left
hand side of
Eq.(\ref{a8}) in the form
\be
\label{a9}
{\cal G}^{(K)}_{i\alpha}(t,t',x')=\int
dt_1\left[\frac{1}{i\partial/\partial t -\hat
H(t)}\right]_{ij}\!\!\!\! (t,t_1)\
W_{j\beta}^\dag{\cal G}_{\alpha\beta}(t_1,t',0,x'),
\ee
The corresponding advance component is zero. Here
$1/(i\partial/\partial t -\hat H(t))$ is the retarded component of
the electron Green's function in the dot. This definition is different
from that given in the main part of the paper, see Eq.(\ref{11}).
The latter will appear naturally in the end of this calculation with
an additional term $\sim W^\dag W$, see \req{14},
describing  escape of electrons from the dot through the leads.
We represent Eq.(\ref{a7}) in the form
\begin{eqnarray}
\fl
{\cal G}^{(K)}_{\alpha\beta}(t,t',x,x') & = & \K_{\alpha\beta}(t-t',x-x')
+ \int dt_1dt_2 \R_{\alpha\gamma}(t-t_1,x)
\nonumber
\\
&&
\times
\left[W\frac{1}{i\partial/\partial t -\hat
H(t)}W^\dag \right]_{\gamma\delta}\!\!\!\!\! (t_1,t_2)
{\cal G}^{(K)}_{\delta\beta}(t_2,t',0,x'),
\label{a10}
\end{eqnarray}
take the limit $x=0$ and, using
$\R_{\alpha\beta}(t-t',0)$ from Eq.(\ref{a4}), obtain
 for $x'<0$
\be
{\cal G}^{(K)}_{\alpha\beta}(t,t',0,x')=
\int\!\! dt_1
\left[1-\hat W \frac{i\pi\nu}{i\partial/\partial t -\hat
H(t)}\hat W^\dag \right]^{-1}_{\alpha\delta}\!\!\!\!\!\! (t,t_1)
G^{(K)}_{\delta\beta}(t_1,t',0,x').
\label{a11}
\ee
Substituting this expression to \req{a10} and taking $x=+|\delta|\to
0$, we find
\be
\label{a12}
{\cal G}^{(K)}_{\alpha\beta}(t,t',+|\delta|,x')=\int\!\! dt_1{
\cal S}_{\alpha\gamma}(t,t_1)\K_{\gamma\beta}(t_1-t',-x'),
\quad  x'<0 ,
\ee
where the scattering matrix ${\cal S}_{\alpha \beta}(t,t')$ is
defined by \req{Sdef}.

Equation (\ref{a12}) is valid for $x'<0$. We have to repeat the procedure
described above to calculate the electron Green's function in the
leads for $x'>0$.
Since the equations which determine evolution of the Green's function
from $x'<0$ to $x'>0$ are conjugated to those for $x$, we obtain
\be
\label{a13}
{\cal G}^{(K)}_{\alpha\beta}(t,t',+|\delta|,+|\delta|)
 =
\int\!\!\int\!\! dt_1dt_2
{\cal S}_{\alpha\gamma}(t,t_1)\K_{\gamma\delta}(t_1-t_2,0)
{\cal S}^\dag_{\delta\beta}(t_2,t').
\ee
The currents in the left (right) leads are given by
\be
\label{a14}
\langle I_{\rm l(r)}(t)\rangle =ev_{\rm F}
\sum\limits_{\alpha \in L(R)}\!\!\left(
{\cal G}^{(K)}_{\alpha\alpha}(t,t,+|\delta|,+|\delta|)-
{\cal G}^{(K)}_{\alpha\alpha}(t,t,-|\delta|,-|\delta|)
\right),
\ee
where $\alpha=1\dots\Nl$ for left lead and $\alpha=\Nl+1\dots\Nch$ for
right lead; coordinate $\delta$ is in the lead just before the
contact with the dot:
$\delta>0$ and $\delta\to 0$. Function
${\cal G}^{(K)}_{\alpha\alpha}(t,t,-|\delta|,-|\delta|)$ is taken for
incoming electrons and is given by Eq.(\ref{a6}) and consequently,
\be
\label{a13'}
{\cal G}^{(K)}_{\alpha\alpha}(t,t,-|\delta|,-|\delta|)=
f(+0), \quad
f(t)=\int_{-\infty}^{+\infty} e^{i\omega t}\tilde
f(\omega) \frac{d\omega}{2\pi}.
\ee

The total dc current from the dot should be zero
$\langle I_{\rm l}\rangle +\langle I_{\rm r}\rangle=0$
to ensure no charge accumulation on the dot,
where
 $\langle I_{\rm l(r)}\rangle = \int_0^{\To}
 \langle I_{\rm l(r)}(t)\rangle dt/\To$. Therefore,
we can
rewrite the expression for dc
current through the dot as
\be
\label{a16}
I(t)=\frac{\Nr \langle I_{\rm l}\rangle-\Nl\langle I_{\rm  r }\rangle}{N_{\rm
ch}}=\langle I_{\rm l }\rangle =-\langle I_{\rm  r }\rangle.
\ee
Substituting Eqs.~(\ref{a13}) and (\ref{a13'}) into Eqs.~(\ref{a14})
and using Eq.~(\ref{a16}) we obtain Eq.~(\ref{10}).


\section{}
\label{app:B}

In this Appendix we derive Eq.~(\ref{4.3}) for the current noise
correlation function through a quantum dot.
The quantum mechanical operator of the current through left
(right) lead is
\be
\label{b.5.1a}
I_{\rm l(r)}(t) =
e v_{\rm F} \sum_{\alpha\in L(R)}
\left(
\psi^\dagger_{\alpha}(t,+\delta) \psi_{\alpha}(t,+\delta) -
\psi^\dagger_{\alpha}(t,-\delta) \psi_{\alpha}(t,-\delta)
\right)
\ee
with
$\psi_{\alpha}(t,\pm \delta)$
being the operator for outgoing ($+\delta$) or incoming ($-\delta$)
electrons through channel $\alpha$, cf. to \req{a14}.

Substituting the expression for the current operator, \req{b.5.1a}
into Eq.~(\ref{4.2}) and using the charge conservation in the dot
on time $\To\gg 1/\omega$, we obtain the following expression for the
current correlation function (below $\delta\to 0$, but $\delta>0$):
\begin{eqnarray}
\nonumber
\fl S  & = &
e^2 v_{\rm F}^2\int_0^{\To} dtdt'
\Big( {\rm tr}
\left\{ \hat \Lambda \hat {\cal G}^<(t',t,+\delta,+\delta)
\hat \Lambda \hat {\cal G}^> (t,t',+\delta,+\delta)\right\}
\\
\fl
&&-
{\rm tr}
\left\{ \hat \Lambda \hat {\cal G}^<(t',t,+\delta,-\delta)
\hat \Lambda \hat {\cal G}^> (t,t',-\delta,+\delta)\right\}
\nonumber
\\
\fl
&&- {\rm tr}
\left\{ \hat \Lambda \hat {\cal G}^<(t',t,-\delta,+\delta)
\hat \Lambda \hat {\cal G}^> (t,t',+\delta,-\delta)\right\}
\nonumber
\\
\fl
&& +
{\rm tr}
\left\{ \hat \Lambda \hat {\cal G}^<(t',t,-\delta,-\delta)
\hat \Lambda \hat {\cal G}^> (t,t',-\delta,-\delta)\right\}
\Big).
\label{b.5.2}
\end{eqnarray}
Here we introduced the electron Green's functions in the leads
according to the following definitions (for a review of the Keldysh
Green's function formalism see \cite{RS}):
\be
\label{b.5.3a}
{\cal G}_{\alpha\beta}^< (t,t',x,x')
=i \langle \psi^\dagger_{\beta}(t',x') \psi_{\alpha}(t,x)
\rangle,\ \
{\cal G}_{\alpha\beta}^>(t,t',x,x')
=
- i \langle \psi_{\alpha}(t,x) \psi^\dagger_{\beta}(t',x') \rangle.
\ee
The Green  functions $\hat {\cal G}^{<,>}$
can be written in
terms of the retarded, advanced and Keldysh Green's functions:
\begin{eqnarray}
\label{b.5.4a}
\fl
\hat {\cal G}^{<}(t,t',x,x') = \frac{1}{2} \left(
\hat {\cal G}^{(K)}(t,t',x,x')-\hat {\cal G}^{(R)}(t,t',x,x')
+ \hat {\cal G}^{(A)}(t,t',x,x')
\right),
\\
\fl
\label{b.5.4b}
\hat {\cal G}^{>}(t,t',x,x') = \frac{1}{2} \left(
\hat {\cal G}^{(K)}(t,t',x,x') + \hat {\cal G}^{(R)}(t,t',x,x')
- \hat {\cal G}^{(A)}(t,t',x,x')
\right).
\end{eqnarray}

The next step is to represent the Green's functions as a product of
incoming electron Green's functions, Eqs.~(\ref{a3})-(\ref{a6}), and
the scattering matrix, \req{Sdef}. The procedure is similar to
one, described in \ref{app:A}. We have the following
relations:
\begin{eqnarray}
\label{b.5.5a}
\fl
\hat {\cal G}^{(R)}(t,t',-\delta,+\delta)
& = &
\hat {\cal G}^{(A)}(t,t',+\delta,-\delta)=0
\\
\fl
\label{b.5.5b}
\hat {\cal G}^{(R,A)}(t,t',+\delta,+\delta)
& = &
\hat G^{(R,A)}(t-t',0),
\\
\fl
\label{b.5.5c}
\hat {\cal G}^{(K)}(t,t',+\delta,+\delta)
& = &
\int
\hat {\cal S}(t,t_1) \hat G^{(K)}(t_1-t_2,0)
\hat {\cal S}^\dagger(t_2,t') dt_1 dt_2,
\\
\fl
\label{b.5.5d}
\hat {\cal G}^{(R,K)}(t,t',+\delta,-\delta)
& = &
\int
\hat {\cal S}(t,t_1) \hat G^{(R,K)}(t_1-t',+\delta)  dt_1 ,
\\
\fl
\label{b.5.5e}
\hat {\cal G}^{(A,K)}(t,t',-\delta,+\delta)
& = &
\int
\hat G^{(A,K)}(t-t_1,-\delta)
\hat {\cal S}^\dagger(t_1,t') dt_1.
\end{eqnarray}

Now the derivation of Eq.~(\ref{4.3}) reduces to simple algebraic
calculations. With the help of \reqs{b.5.4a} and
\rref{b.5.4b} we rewrite Eq.~(\ref{b.5.2}) in terms of the retarded,
advanced and Keldysh components of the Green  function. Then we represent
these components
as a product of scattering matrices and the Green's functions
of the incoming electrons, using Eqs.~(\ref{b.5.5a})-(\ref{b.5.5e}).
The result is given by Eq.~(\ref{4.3}).

\end{document}